\documentclass[useAMS,usenatbib]{mn2e}
\usepackage{graphicx}
\usepackage{graphics}
\usepackage{epstopdf}
\usepackage{float}
\usepackage{bm}
\usepackage{epsf}
\usepackage{url}
\usepackage{epsfig}
\usepackage{natbib}
\usepackage{amsmath}

\newcommand{\msol}{\,$h^{-1}$ M$_{\odot}$}

\title[Sub-Megaparsec Individual Photometric Redshift]{Sub-Megaparsec Individual Photometric Redshift Estimation from Cosmic Web Constraints}
\author[Aragon-Calvo M.A. et al.]{M.A. Aragon-Calvo$^{1}$\thanks{E-mail:maragon@ucr.edu}, Rien van de Weygaert$^{2}$, Bernard J.T. Jones$^{2}$, Bahram Mobasher$^{1}$\\
$^{1}$Department of Physics and Astronomy. University of California, Riverside, CA, USA.\\
$^{2}$Kapteyn Astronomical Institute, University of Groningen, P.O. Box 800, 9700 AV, Groningen, The Netherlands.\\}
\begin{document}

\date{Submitted to MNRAS}

\pagerange{\pageref{firstpage}--\pageref{lastpage}} \pubyear{2002}
\maketitle
\label{firstpage}

\begin{abstract}

We present a method, PhotoWeb, for estimating photometric redshifts of \textit{individual galaxies}, and their equivalent distance, with megaparsec and even sub-megaparsec accuracy using the Cosmic Web as a constraint over photo-$z$ estimates. PhotoWeb redshift errors for individual galaxies are of the order of $\Delta z \simeq 0.0007$, compared to errors of $\Delta z \simeq 0.02$ for current photo-$z$ techniques. The mean redshift error is of the order of $5\times10^{-5}-5\times10^{-4}$ compared to mean errors in the range $\Delta z \simeq 0.001-0.01$ for the best available photo-$z$ estimates in the literature. Current photo-$z$ techniques based on the spectral energy distribution of galaxies and projected clustering produce redshift estimates with large errors due to the poor constraining power the galaxy's spectral energy distribution and projected clustering can provide.  The Cosmic Web, on the other hand, provides the strongest constraints on the position of galaxies. The network of walls, filaments and voids occupy $\sim \%10$ of the volume of the Universe, yet they contain $\sim \%95$ of galaxies. The cosmic web, being a cellular system with well-defined boundaries, sets a restricted set of intermittent positions a galaxy can occupy along a given line-of-sight. 
Using the information in the density field computed from spectroscopic redshifts we can narrow the possible locations of a given galaxy along the line of sight from a single broad probability distribution (from photo-$z$) to one or a few narrow peaks. Our first results improve previous photo-$z$ errors by more than one order of magnitude allowing sub-megaparsec errors in some cases. Such accurate estimates for tens of millions of galaxies will allow unprecedented galaxy-LSS studies. In this work we apply our technique to the SDSS photo-$z$ galaxy sample and discuss its performance and future improvements. 
\end{abstract}

\section{Introduction}

Spectroscopic redshifts are our main tool for measuring accurate distances to galaxies and studying the large scale structure in the universe. They are however expensive to measure due to current technological limitations, and as a result, only a small fraction of the galaxies observed in photometry are able to be observed in spectroscopy. Photometric redshifts, on the other hand, offer a cheap alternative to estimate redshifts of a large number of galaxies using broadband filters, at the expense of having large errors (tens of megaparsecs). Photo-$z$ techniques perform, at their core, some sort of matching or interpolation between the observed Spectral Energy Distribution (SED) of a galaxy and a known observed or model template  (\citet{Connolly95,Benitez00,Mobasher07,Csabai07,Cunha09,Carliles10}, see also \citet{Hildebrandt10} for an extensive review and comparison between different photo-$z$ methods). A notable exception to these conventional methods is the approach of \citet{Kovac10} in which the photo-$z$ probability density is modified by using the local density as a constraint, reducing the photo-$z$ errors to within the scale of the smoothing kernel used to probe the density field. Standard photo-$z$ techniques compute redshift estimates for each galaxy independently taking into account only their SED and have typical redshift estimates with errors of $\Delta z\sim 0.01-0.02$, equivalent to $\sim 40-80$ Mpc. A direct way of reducing the redshift errors is by increasing the number of observed color bands (e.g. the J-PAS multi band survey with more than 50 narrow bands). Such multi-band surveys can be regarded as low-resolution spectroscopy but still produce errors of the order of $\Delta z=0.003(1+z)$. Recent improvements on photo-$z$ estimations include clustering information from a reference sample of galaxies for which accurate redshift estimations are available, typically from spectroscopy. This is done by means of correlation functions \citep{Schneider06, Newman08, Jasche12, Menard13, Rahman14} achieving \textit{mean redshift errors} of the order of $\Delta z = 0.001-0.01$ but still providing large errors for \textit{individual galaxies}. While the two-point correlation function provides a measure of clustering, it cannot fully describe the highly anisotropic distribution of galaxies. For this, high-order correlation functions are required \citep{Fry84,Szapudi91}, and even then still may miss critical phase information, as high-order correlation functions do not capture all clustering information in a sufficiently non-Gaussian field \citep{Carron11,Carron12}. Correlation functions, while providing some indication of structure, are effectively blind to the cellular nature of the large-scale galaxy distribution \citep{Zeldovich82, Aragon13a} which sets the strongest constraints on the position of a galaxy along the line-of-sight (LoS).


%
\begin{figure}
  \begin{center}
       \includegraphics[width=0.4\textwidth,angle=0.0]{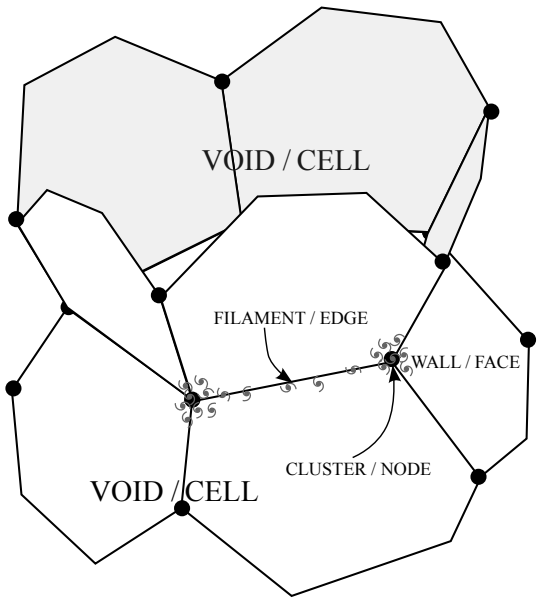}
  \end{center}
  \vspace{-0.5cm}
  \caption{The distribution of galaxies as a cellular system. Voids, walls, filaments and clusters delineated by luminous galaxies correspond to cells, faces, edges and nodes of a cellular system. For illustration purposes we draw several galaxies delineating two nodes (clusters) and their common edge (filament).}
  \label{fig:void-cell}
\end{figure}

%
\subsection{Constraints from the Cosmic Web} \label{sec:cosmic_web_constraints}

The distribution of galaxies forms a cellular system in which large empty voids (cells) are surrounded by a network of increasingly denser walls (faces), filaments (edges) and clusters (nodes) as illustrated in Fig. \ref{fig:void-cell}. The origin of the cellular nature of the Universe can be understood by the sequential anisotropic gravitational collapse of a primordial three-dimensional cloud into two-dimensional walls, one-dimensional filaments and finally dense compact clusters \citep{Zeldovich70}. How the individual structural elements of the large scale structure are arranged is (partially) described by the \textit{cosmic web theory} \citep{Bond96} which predicts the emergence of filaments in the space in-between adjacent peaks in the density field  corresponding to clusters, and in general, the emergence of a network of filaments joining clusters, groups and galaxies. Another useful picture for understanding the emergence of cellular patterns in the distribution of galaxies is the one given by the \textit{Bubble Theorem} \citep{Icke84} in which voids act as expanding bubbles that intersect as they fill the available space, forming the dense ridges with walls, filaments and clusters.

Voids occupy most of the volume of the Universe ($\sim 90\%$) while containing only a minor fraction of the observed luminous galaxies ($\sim 5\%$). This means that walls, filaments and clusters, accounting for $\sim10\%$ of the volume in the Universe, hold 95\% of all the galaxies \citep{Aragon10,Cautun14}.  This asymmetry between volume occupancy and galaxy number density in addition to the cellular character of the galaxy distribution set strong constraints on the positions a galaxy can occupy given its surrounding matter configuration. The density field sampled along a LoS is characterized by a set of narrow, intermittent, semi-periodic peaks \citep{Kirshner81,Broadhurst90,Weygaert91} separated by shallow valleys with a range of sizes delimited by the maximum void size of the order of $\sim 20-50$Mpc \citep{Platen08}.
We use this information to narrow the possible locations of a given galaxy along the LoS from a broad probability distribution (from photo-z) to a few narrow peaks given by the structures in the cosmic web.

%
\begin{figure*}
  \begin{center}
       \includegraphics[width=0.99\textwidth,angle=0.0]{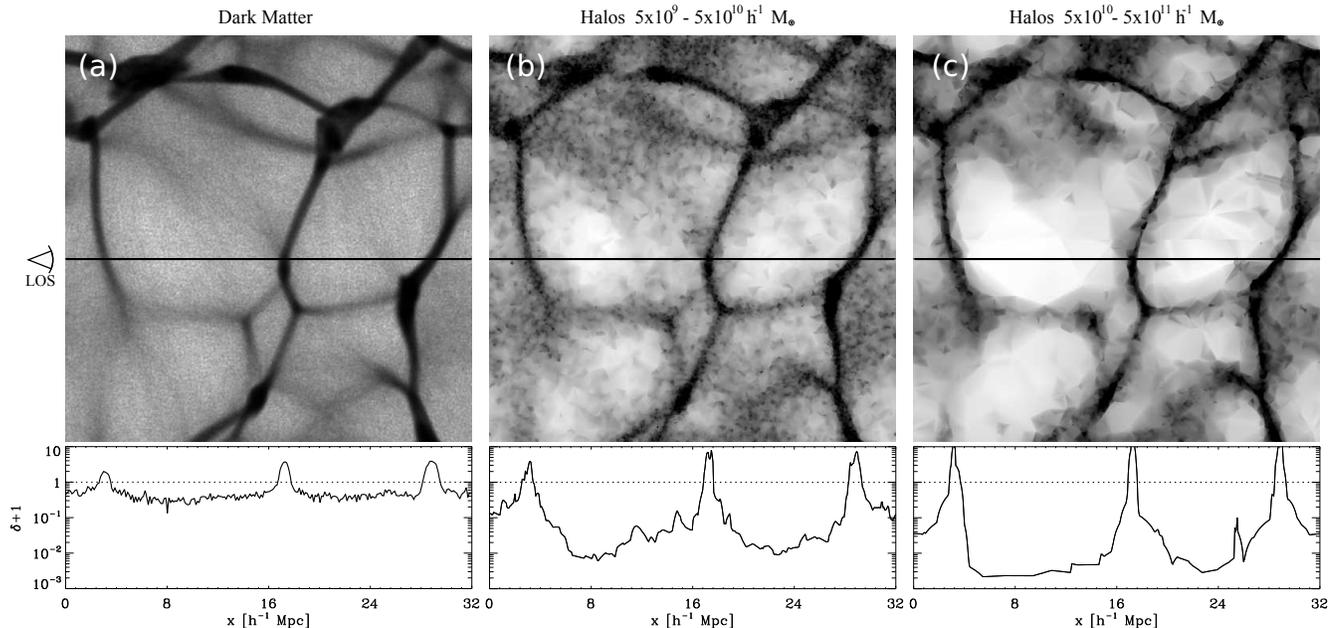}
  \end{center}
  \vspace{-0.5cm}
  \caption{Halo occupation constraints in the cosmic web. Halo bias has a sharpening effect in the features of the cosmic web. a) dark matter ensemble density field, b) low-mass halos ($5\times10^{9} <$ M $<$ $5\times10^{10}$ \msol) ensemble  density field, c) high-mass halos ($5\times10^{10} <$ M $<$ $5\times10^{11}$ \msol) density field. The density fields were computed by stacking 256 realizations from the \textit{Multum In Parvo} (MIP) correlated ensemble \citep{Aragon13b}. The lower panels show the density profile for the LoS indicated on the left side of the top panels.  Halo bias produces an excess of haloes in dense regions and increasingly empty voids for increasing halo mass. Haloes more massive than $10^{11}$ \msol are practically absent from voids. Note that galaxies occupy narrow peaks along the LoS and this becomes more prominent with increasing halo mass. The ensemble density field is equivalent to observing 256 independent lines of sight so the lack of massive haloes is a real effect and not the product of sampling noise.}
  \label{fig:MIP_bias}
\end{figure*}

%
\subsection{Bias-enhanced cosmic structures}

Galaxies are \textit{biased} tracers of the underlying matter distribution \citep{Kaiser84,Mo96}. While the matter density field is a volume-filling continuous field, galaxies are concentrated along the dense ridges of the density field and avoid under-dense environments. The galaxy population can be considered as a stochastic sampling of  a continuous locally-biased matter density field. The local bias function generally produces an over-concentration (relative to the matter density field) of galaxies in dense regions (walls, filaments and clusters), and an exponential suppression within voids as their halo mass function is shifted towards low-masses \citep{Gottlober03,Goldberg04,Neyrinck14}. This is a major reason why haloes are so rare in voids, and why locating galaxies along the cosmic web outlining voids is a good approximation. Figure \ref{fig:MIP_bias} shows the local biasing of dark matter haloes as a function of density and halo mass. The cosmic web features delineated by massive haloes are sharp, unlike the smoother structures delineated by less massive haloes and the even more shallow profile of the dark matter. If we consider the density field as a probability density distribution indicating the regions where it is more likely to find a galaxy, it is then clear that the biased galaxy distribution provides strong constraints on the location of a galaxy along the LoS. The constraints are more significant (the difference between peaks and valleys increases) for increasingly massive galaxies as indicated in Fig. \ref{fig:MIP_bias}.

%
\subsection{The dataset}

The analysis presented in this work is based on the Sloan Digital Sky Survey (SDSS, \citet{York00,Abazajian09}), containing more than 1 million galaxies with measured spectroscopic redshift and $\sim200$ million galaxies with photometric redshift computed using a KD-tree nearest neighbor algorithm \citep{Csabai07}. We restrict our analysis to galaxies with photo-$z$ uncertainties below $\Delta z =0.015$, giving a total of $\sim6\times10^6$ galaxies with good photometric redshifts. The spectroscopic sample was used to compute density fields as described below and as a reference for error estimation for both photo-$z$ and the method presented here (see  \ref{app:sql} for details in the sample selection).


\section{The PhotoWeb method} \label{sec:PhotoWeb}

Here we provide an outline of our redshift measurement technique, hereafter referred to as PhotoWeb. More detailed description of the method will be presented in sections \ref{sec:DTFE} to \ref{sec:assign_photoweb}. Conceptually, PhotoWeb is based on two key observations:

\begin{itemize}
\item[i)]  The distribution of galaxies follows a well-defined cellular structure with large empty regions surrounded by denser walls, filaments and clusters
\item[ii)] Galaxies are a biased sampling of the underlying matter density field, over-populating the narrow high-density ridges of the cosmic web while avoiding the vast under-dense voids.
\end{itemize} 

\noindent These two properties of the Universe alone set strong constraints on the most probable location a galaxy can occupy. We should be able to predict the most likely position of a galaxy along the LoS from the distribution of the surrounding galaxies by combining the photo-$z$ PDF $P(z)_{\textrm{\tiny{photo}}}$, which depends on the particular method used, with constraints derived from the cosmic web itself.

The main requirement for PhotoWeb is a reference sample of galaxies with accurate redshifts \footnote{In practice this corresponds to spectroscopic redshifts but it can be any other method}. This reference sample must be dense enough to resolve the network of walls, filaments and clusters surrounding voids, since these are the structures that will be used to constrain the location of galaxies along the LoS. Strictly speaking we only need to resolve individual voids since their boundaries give the locations of walls, filaments and clusters. Finding voids is relatively easy using methods that probe the topology of the density field as described below. This means that we could, in principle, use a sample of photometric redshifts as the reference sample provided their typical redshift uncertainty is smaller than the mean void size. In this work we will use the SDSS spectroscopic sample since the mean photo-$z$ uncertainties are of the order of $\sim 80$Mpc which is several times the mean void size.

The cosmic structures in the reference sample are then used to constrain the position of a sample of target galaxies for which we have a redshift probability density distribution with a large uncertainty, in our case given by a photo-$z$ method, $P(z)_{\textrm{\tiny{photo}}}$. The photo-$z$ probability distributions have in most cases a dominant peak with a dispersion of the order of $\sigma_{\textrm{\tiny{photo}}} \sim 0.01-0.02$ (for the SDSS). On the other hand, the probability density function computed from the density field (along the LoS) $P(z)_{\textrm{\tiny{den}}}$, is a complex function with multiple wide-spaced narrow peaks, each with individual dispersions of megaparsec/sub-megaparsec scale. The key idea in PhotoWeb is combining the broad but unique solution provided by photo-$z$ with the narrow but degenerate constraint provided by the cosmic web in order to collapse the redshift probability distribution into, ideally, one single narrow peak. 

Our goal is to compute the probability density distribution given by the cosmic web for a given LoS. In practice we divide this function into two components, one that depends on the local density, which dominates the PDF and one that depends on the geometry of the cosmic web, which adds minor corrections to special cases as explained in the following sections. The photoweb probability density distribution is then:

\begin{equation}\label{eq:P_web}
P(z)_{\textrm{\tiny{ web}}} = P(z)_{\textrm{\tiny{photo}}}  \cdot P(z)_{\textrm{\tiny{den}}}  \cdot P(z)_{\textrm{\tiny{geo}}}.
\end{equation}

\noindent Where $P(z)_{\textrm{\tiny{den}}}$ is the density field sampled along the LoS and $P(z)_{\textrm{\tiny{geo}}}$ is a function that follows the geometry of the cosmic web in a similar way to the density field but giving equal weight to all cosmic structures. The normalization of $P(z)_{\textrm{\tiny{ web}}}$ is not crucial for our purposes as explained in the following sections. In what follows we provide a detailed description of the PhotoWeb method pipeline. We describe the present state of the method, currently in active development and discuss, when appropriate, possible extensions or optimizations to be applied in future versions.

%
\begin{figure}
  \begin{center}
       \includegraphics[width=0.49\textwidth,angle=0.0]{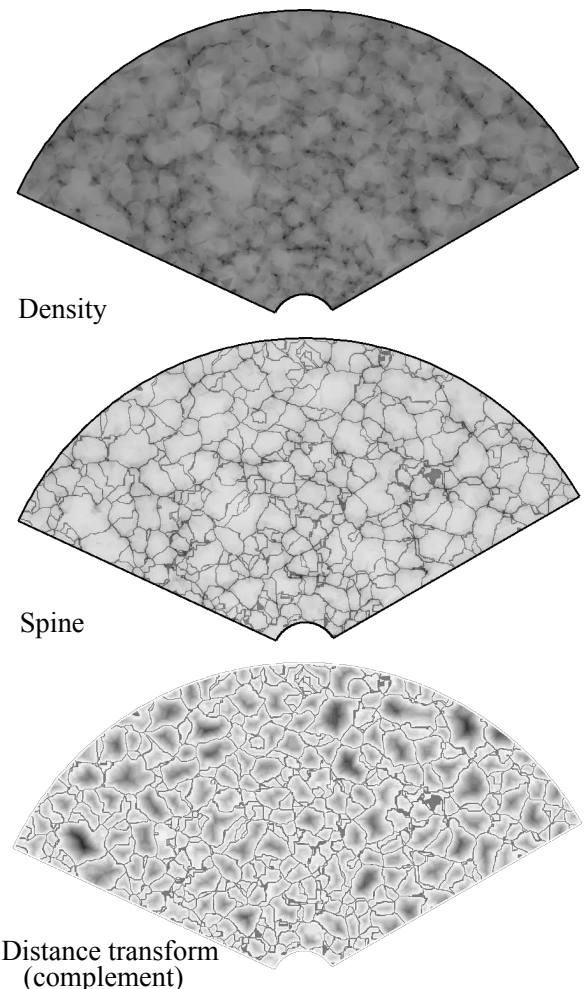}
  \end{center}
  \vspace{-0.5cm}
  \caption{From top to bottom: density field, spine (walls, filaments and clusters) and distance transform on a thin slice computed from the SDSS spectroscopic  sample. For clarity the density field is inverted so dark pixels indicate high-density regions. The spine in the middle panel is shown as the dark contours delineating voids. The distance transform (also inverted) has similar properties as the density field, decreasing from the spine into the centers of voids.}
  \label{fig:den-spi-dis}
\end{figure}

%
\subsection{Density field constraints, $P(z)_{\textrm{\tiny{den}}}$}\label{sec:DTFE}

The reconstruction of the underlying continuous density field from a discrete galaxy distribution is one of the key steps in the PhotoWeb method. The fact that we need a pre-existing density field limits the applicability of PhotoWeb to photometric surveys with overlapping spectroscopy coverage sufficiently dense to sample the structures in the cosmic web. The SDSS spectroscopic sample used in this work provides good spectroscopic coverage up to $z < 0.12$ (see Appendix \ref{app:selection_function}). We begin by converting spectroscopic redshifts $z{\textrm{\tiny{spec}}}$ to distances using the approximation:

\begin{equation}\label{eq:red_dista}
d = z{\textrm{\tiny{spec}}} \; c / H_0
\end{equation}

\noindent where $c$ is the speed of light and $H_0 = 73$km s$^{-1}$Mpc$^{-1}$ is the Hubble's constant. This approximation is, for our purposes, sufficient to compute distances up to $z < 0.12$. 

\begin{figure*}
  \centering
      \includegraphics[width=0.95\textwidth,angle=0.0]{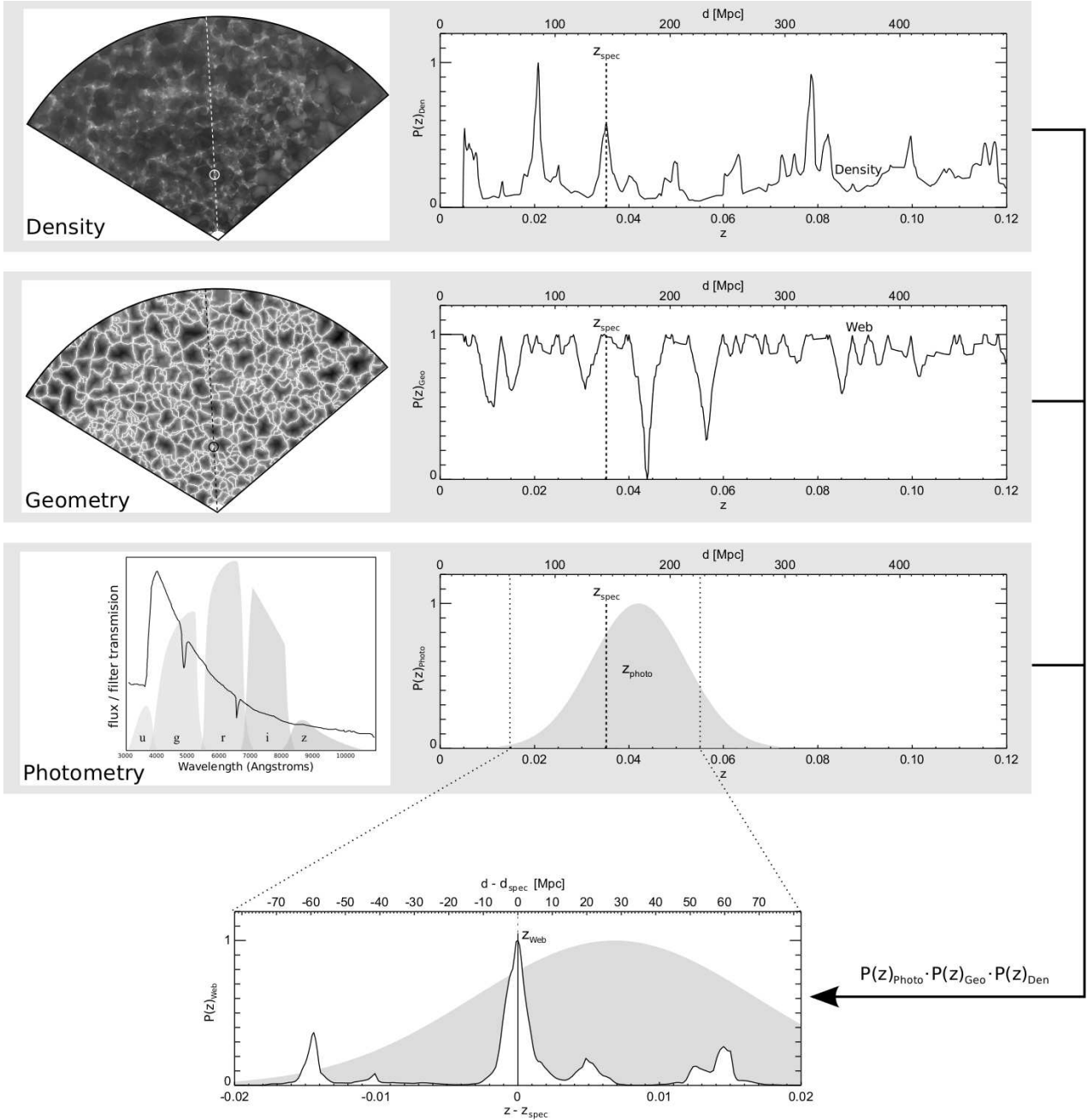}
  \caption{ {\bf Computing the PhotoWeb PDF ($P(z)_{\textrm{\tiny{web}}}$)}. This diagram corresponds to a galaxy with spectroscopic redshift of $z=0.035$. We show the three components involved in the computation of $P(z)_{\textrm{\tiny{web}}}$: density ($P(z)_{\textrm{\tiny{den}}}$ (top), geometry $P(z)_{\textrm{\tiny{geo}}}$ (middle) and photometry $P(z)_{\textrm{\tiny{photo}}}$ (bottom). The density field and the derived distance transform where computed from the spectroscopic sample. For simplicity we assume $P(z)_{\textrm{\tiny{photo}}}$ to be a Gaussian distribution with dispersion equal to the zErr value (see Appendix \ref{app:sql}) given by the KD-tree photo-$z$ method \citep{Csabai07}. The left panels show the density field, distance transform and SED (the SED does not correspond to this particular galaxy and is used only as an illustration). The panels on the right side show the corresponding PDF normalized to 1 at the maximum. The panel at the bottom shows the product of the three PDFs and the resulting $P(z)_{\textrm{\tiny{web}}}$. While the photometric redshift has a distance error of $\Delta z_{\textrm{\tiny{photo}}}\sim 0.02$ equivalent to $\sim 30$Mpc, the assigned PhotoWeb redshift is, in this case, indistinguishable from its spectroscopic counterpart.}
   \label{fig:photoweb_diagram}
\end{figure*} 

%
\subsubsection{Removing redshift artifacts}

When converting spectroscopic redshifts to distances one must be aware of artifacts arising from peculiar velocities of galaxies in dense environments, where small-scale non-linear interactions produce large velocity dispersions. These \textit{redshift distortions} smear compact dense groups and clusters of galaxies along the LoS,  producing the so-called Fingers-of-God (FoG, \citep{Kaiser87}). We identify and correct FoG using the approach based on a Friends of Friends groups finder with an elliptical search radius introduced by \citet{Tegmark04}. Groups elongated along the LoS are identified and compressed such that their dispersion along the LoS is equal to the dispersion in the plane of the sky (see Appendix \ref{sec:fog_compression} for details). Another redshift distortion is the \textit{Kaiser effect} \citep{Kaiser87} originating from large-scale peculiar velocities that compress low-density structures along the LoS. We do not attempt to correct for this effect as it is small and can even enhance cosmic structures in contrast to the FoG that smear groups along the LoS.

%
\subsubsection{Density field estimation and cosmic web reconstruction}

From the (FoG-corrected) galaxy distribution we reconstruct the underlying density field using the Delaunay Tessellation Field Estimator (DTFE) technique (\citet{Schaap00}, see also \citet{Platen11}). Given a spatial distribution of points (galaxies in our case), DTFE estimates the density at the position of each galaxy as the inverse volume of the adjacent Delaunay tetrahedra, i.e.  its contiguous Voronoi cell. The density field is then linearly interpolated inside the tetrahedra to produce a continuous density field on a sampling grid. The Delaunay tessellation follows the anisotropic features in the galaxy distribution and propagates local density estimates across the spaces between sampling points, providing a first-order \textit{structure interpolation} scheme (Aragon et al. in preparation, see Appendix \ref{app:delaunay_reconstruction}). In order to compensate for the decrease in the galaxy number density with distance we weight each individual density estimate with the galaxy selection function (see Appendix \ref{app:selection_function}). For each galaxy the density estimation is then given by

\begin{equation}
f(\mathbf{x_i}) =  \frac{(1+D)}{ \psi(z_i) V(\mathcal{W}_i)}.
\end{equation}

\noindent Where $D=3$ is the number of spatial dimensions in which the point distribution is embedded, $V(\mathcal{W}_i)$ is the volume of the contiguous Voronoi cell of the point $i$ and $\psi(z_i)$ is the selection function defined as:

\begin{equation}
\psi(z_i) = e^{-(z_i/z_r )^{\beta}}.
\end{equation}

\noindent Where $z_r$ is a characteristic redshift defining the peak in the redshift distribution. A single density determination at the center of a voxel tends to introduce aliasing artifacts in dense regions (where the mean inter-galaxy separation can be much smaller than the distance between sampling points. Aliasing artifacts in the density field manifest as hot voxels, which can affect, among other things, the topological properties of the density field. In order to reduce aliasing artifacts we super-sample each voxel with a sub-grid with three times higher resolution and set the density at each voxel as the mean of the 27 sub-grid locations within the voxel. We use a fast and efficient implementation of the DTFE algorithm based on the publicly available CGAL library\footnote{www.cgal.org}. Computing the density field on a $1024\times512\times512$ grid using DTFE takes a few minutes on a modern workstation.

We then use the density field sampled along the LoS of the target galaxy directly as the the probability of a galaxy residing at redshift $z$ along the LoS, $P(z)_{\textrm{\tiny{den}}}$. 


%
\subsection{LSS geometry constraints, $P(z)_{\textrm{\tiny{geo}}}$}

The density field reconstructed from the distribution of galaxies has a wide dynamic range, as illustrated in Fig. \ref{fig:MIP_bias}. The difference in density between the interior of a void and the center of a cluster or filament is of several orders of magnitude. This property of the galaxy density field can introduce errors in the redshift estimation in cases when the LoS of a galaxy intersects a filament or wall in the vicinity of a cluster, or when the LoS intersects a cluster within a couple of standard deviations from the mean in the photo-$z$ probability distribution of the target galaxy. In these situations $P(z)_{\textrm{\tiny{den}}}$ is dominated by the cluster, even if the cluster is relatively far from the true position of the galaxy. In order to compensate for this effect, we include an additional constraint that determines the likelihood of a galaxy at a given position along the LoS based on the \textit{geometry} of the cosmic web and not on its density. Since we are only interested in the geometry of the structures, we use a transformation that gives equal weight to all cosmic structures (here and in what follows, the term \textit{cosmic structure} refers to the set of walls, filaments and clusters and excludes voids which are their complement). We use the distance from a target galaxy to the nearest cosmic structure to estimate the probability of the galaxy being associated to the nearest cosmic structure. This is done by means of the \textit{distance transform}. The distance transform is a mathematical morphology operator, usually applied to a binary image, that specifies the distance from each voxel to the nearest non-zero voxel (or boundary) according to a given metric. The distance transform of a set of points on a grid $P \subseteq G$ is computed by associating to each grid location the distance to the nearest point in $P$ as:

\begin{equation}\label{eq:distance_transform}
D_P(p) = \min_{q \in P} d(p,q)
\end{equation}

\noindent where $d(p,q)$ is the euclidean distance between $p$ and $q$. If the target galaxy is located inside a cluster, the result from the density constraint alone is not affected by the geometric constraint. If however, the target galaxy is located inside a filament or wall and there is a nearby cluster, as in the situation described above, then the distance transform will give more weight to the cosmic structure closest to the target galaxy instead of the cluster, providing a way to compensate for the large dynamic range of the densities in the cosmic web.

%
\subsubsection{Cosmic Web identification} \label{sec:SPINE}

In order to compute the distance transform we need a discrete representation of the density field that encodes the cosmic structures as a binary set such as:

\begin{equation}
  M = \left\{
    \begin{array}{cl}
      1 & \textrm{cluster, filament, wall}\\
      0 & \textrm{void}
    \end{array} \right.
\end{equation}

\noindent Which requires a way of separating, or segmenting, voids from the other cosmic structures. We perform a full identification of voids, walls, filaments and clusters in the density field using the SPINE method \citep{Aragon10} which provides, from a continuous scalar field, a discrete label for each voxel in the density field as void, wall, filament or cluster. The SPINE method assigns morphological labels based on the topology of the density field, encoded in the \textit{watershed transform} \citep{Beucher79,Platen07}. Figure \ref{fig:den-spi-dis} shows the SPINE field (middle panel) computed from the density field on the top panel. The SPINE method provides a cellular segmentation of space corresponding to voids, as well as a full description of the cosmic web. In the present work we only use the spine i.e. the union of the set of walls, filaments and clusters as the basis to compute the distance transform.

From the cellular segmentation provided by SPINE we then proceed to compute the distance transform using an efficient algorithm that takes advantage of the space-partitioning character of the watershed transform (see Appendix \ref{app:distance_transform}), giving us an objective way to quantify association to a cosmic web structure. In practice we use the complement of the distance field to produce a monotonically increasing scalar field akin to the density field but giving equal weight to clusters, filaments and walls. The distance transform along the LoS of a galaxy can then be considered a probability density distribution $P(z)_{\textrm{\tiny{geo}}}$ giving us the likelihood of finding a galaxy along the LoS on the basis of the geometry of the cosmic web (see Fig. \ref{fig:den-spi-dis}).

%
\begin{figure*}
  \begin{center}
      \includegraphics[width=0.49\textwidth,angle=0.0]{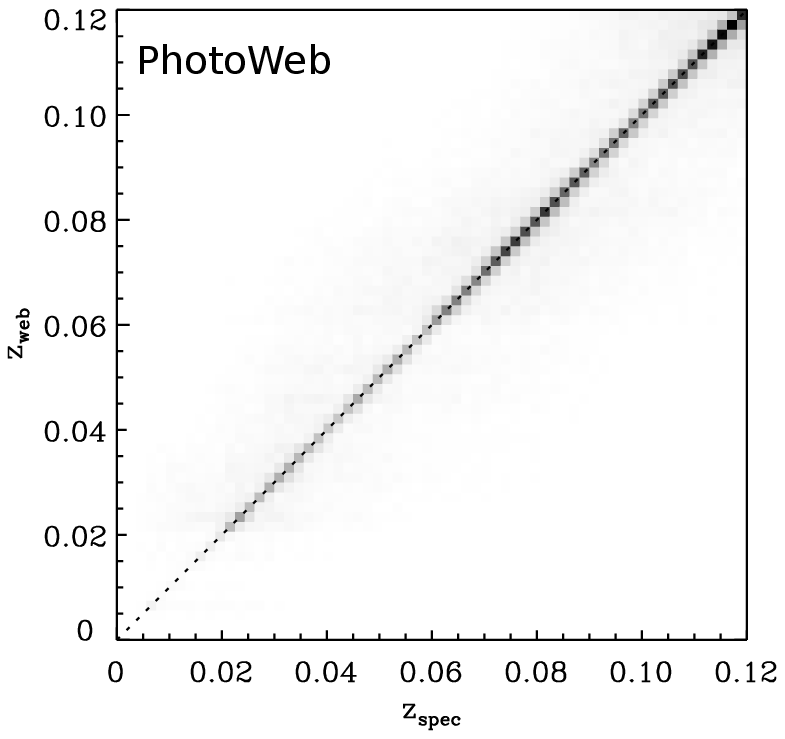}
      \includegraphics[width=0.49\textwidth,angle=0.0]{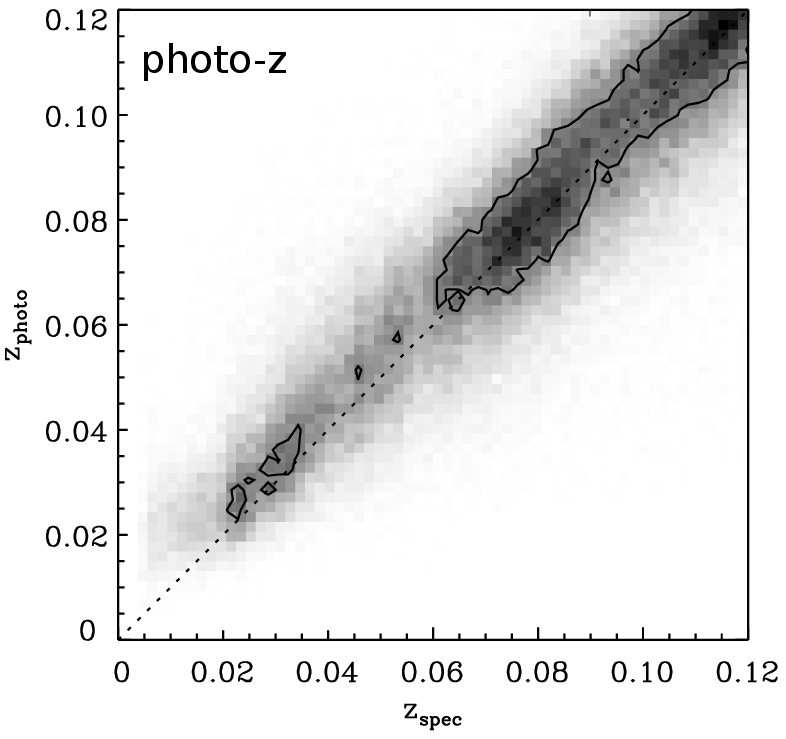}
  \end{center}
  \vspace{-0.5cm}
  \caption{\small {\bf PhotoWeb redshift estimates vs. spectroscopic redshifts.} The redshift range is divided into 32 bins. Gray levels indicate the counts inside each bin. For comparison we show the photometric redshift vs. spectroscopic redshift. }
  \label{fig:photoweb_vs_spec_2d_histogram}
\end{figure*}

%
\subsection{Adding density and cosmic web constraints}

The final step in the PhotoWeb pipeline is computing the product of the photometric probability density distribution and the density and cosmic web constraints as indicated in Equation \ref{eq:P_web}, here repeated for convenience:

\begin{align*}
P(z)_{\textrm{\tiny{ web}}} = P(z)_{\textrm{\tiny{photo}}}  \cdot P(z)_{\textrm{\tiny{den}}}  \cdot P(z)_{\textrm{\tiny{geo}}}  \nonumber
\end{align*}
 
\noindent In practice, for simplicity we replace the full photo-z probability distribution of a given galaxy for a Gaussian distribution centered at the mean redshift and with the dispersion equal to the uncertainly $z_{err}$ provided by the KD-tree photo-$z$ method \citep{Csabai07}. This is a good approximation for most cases where there is a single dominant peak in the photo-$z$ probability distribution. Figure \ref{fig:photoweb_diagram} shows a schematic of the PhotoWeb pipeline starting from the density field reconstruction. The density field probability distribution, $P(z)_{\textrm{\tiny{den}}}$,  is characterized by a few narrow high peaks. Note that the dominant peak in $P(z)_{\textrm{\tiny{den}}}$ is not the peak corresponding to the target galaxy $z=0.035$. There is one significantly higher peak at $z\sim0.02$. If the photo-$z$ probability distribution was broader then the larger peak would likely be selected as the photoweb-$z$. This highlights the importance of having accurate photo-$z$ with narrow PDE. We will discuss this in detail in Section \ref{sec:errors}. In contrast, the features in $P(z)_{\textrm{\tiny{geo}}}$ delineate cosmic structures and are also correlated to the peaks in the density field, but the peaks in $P(z)_{\textrm{\tiny{geo}}}$ have all the same height, thus giving equal significance to all cosmic web structures regardless of their local density. By multiplying the broad single-peak photo-z distribution with the narrow multiple-peak density and distance transform distributions we collapse the probability distribution and obtain a unique solution given by the highest peak in the $P(z)_{\textrm{\tiny{web}}}$ distribution. 

%
\begin{figure*}
  \begin{center}
       \includegraphics[width=0.99\textwidth,angle=0.0]{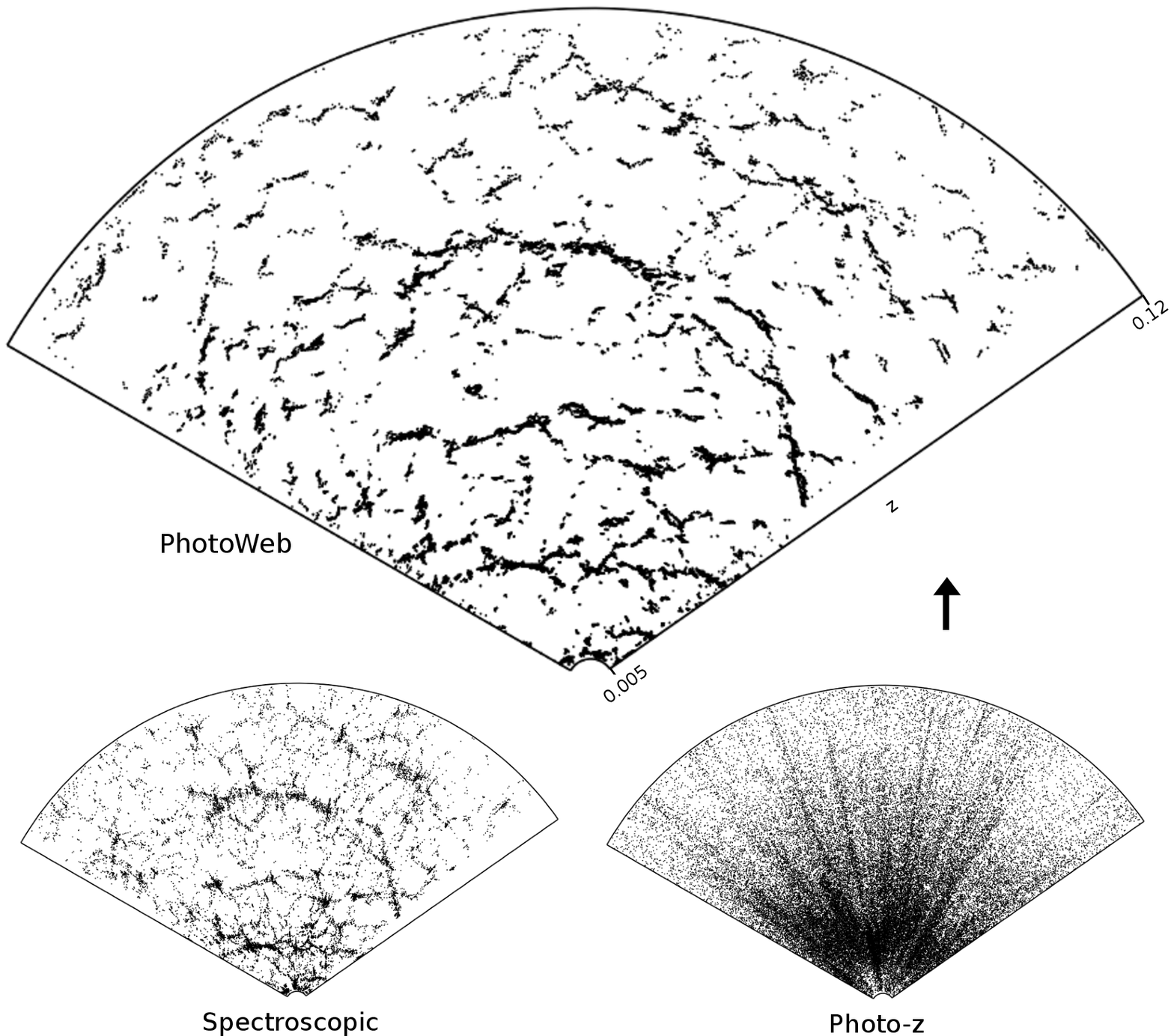}
  \end{center}
  \vspace{-0.5cm}
  \caption{\small {\bf Galaxy distribution from  PhotoWeb, photo-$z$ and spectroscopic redshifts.} We show galaxies in a 10 Mpc-thick slice across the SDSS survey. The bottom-left panel shows the galaxy positions from spectroscopic redshifts without FoG compression. The top-left panel shows the photo-$z$ positions. The top-right panel shows the PhotoWeb positions. The distribution of galaxies from PhotoWeb follow the cosmic web delineated by the spectroscopic reference sample. Note that the distribution of galaxies from photo-$z$ show barely any feature. The PhotoWeb positions are highly biased, there are no galaxies inside the voids and the cosmic structures have very sharp boundaries. There is also a lack of filaments aligned with the LoS as a result of the FoG compression used to derive the density field. Nevertheless the PhotoWeb positions are sticking. This diagram highlights the unique potential of PhotoWeb for small-scale environmental studies.}
  \label{fig:PhotoWeb_SDSS}
\end{figure*}

%
\subsection{Assigning PhotoWeb redshifts}\label{sec:assign_photoweb}

In the present implementation of PhotoWeb we use a simple approach and assign the PhotoWeb -$z$ to the redshift where the distribution $P(z)_{\textrm{\tiny{web}}}$ has its maximum value. This prescription is sufficient for our purposes in this introductory paper, but it also produces artificially large biasing. Since we assign galaxies only to the peak in $P(z)_{\textrm{\tiny{web}}}$ we impose a complete lack of galaxies inside voids. If a target galaxy is located inside a void (which in itself has a very low likelihood) it will be artificially shifted towards the nearest cosmic structure. In terms of the global population, galaxies in low-density environments represent a minor fraction but nevertheless an important one.  One way to alleviate this would be to implement a montecarlo sampling of $P(z)_{\textrm{\tiny{web}}}$ or some transformation of it such as $\log P(z)_{\textrm{\tiny{web}}}$ which provides a more natural density variable in the non-linear regime\citep{Coles91,Neyrinck09,Carron11}. The sampling can also include a biasing prescription matched to the observed local galaxy bias. 

\section{Results}

In this section we present our first results and perform tests in order to understand the errors and systematics in the PhotoWeb method. In order to measure redshift errors we applied PhotoWeb to galaxies in the SDSS spectroscopic sample, which also have photo-$z$ estimates \citep{Csabai07} (see Appendix \ref{app:sql} for details). We use spectroscopic redshifts as a reference, assuming that they provide accurate redshifts and estimate distances, which is valid for low and medium-density environments and breaks down in dense clusters. However since we also correct FoG we simply assume all spectroscopic redshifts and positions to be correct. 

%
\subsection{PhotoWeb vs. photo-$z$ redshift estimation}

We start by comparing how well the PhotoWeb redshifts follow the reference spectroscopic redshift. Figure \ref{fig:photoweb_vs_spec_2d_histogram} shows a  comparison between photoweb redshifts $z_{\textrm{\tiny{web}}}$ and the spectroscopic redshifts $z_{\textrm{\tiny{spec}}}$ used as reference. PhotoWeb redshifts are strongly correlated with the spectroscopic redshifts. The dispersion around the slope unit line in Fig \ref{fig:photoweb_vs_spec_2d_histogram} $\Delta z \sim 0.0006$ equivalent to $\sim 2$Mpc and the FWHM is of the order of $\Delta z \sim 0.0002$ or $\sim 1$Mpc. The 2D distribution is dominated by a narrow diagonal ridge and shallow broad tails. However, the distribution is non-Gaussian and the tails extend beyond what we expect for a Gaussian distribution. As a result of this, the total dispersion is higher than the one estimated from the narrow peak (this is explored in detail in Sec. \ref{sec:error_model}). Nevertheless more than half of the galaxies fall inside the narrow ridge. For comparison we show the same plot for photo-$z$ estimates (right panel in Fig. \ref{fig:photoweb_vs_spec_2d_histogram}). Here the correlation is not tight and the dispersion is significantly larger, of the order of $\Delta z \simeq 0.01-0.02$, corresponding to $\sim40-80$Mpc. The correlation between $z_{\textrm{\tiny{web}}}$ and $z_{\textrm{\tiny{spec}}}$ is very tight even at low redshifts ($z < 0.05$) where the photo-$z$ estimates, used as starting point for PhotoWeb, have consistently higher values than their corresponding spectroscopic redshift. 

%
\subsection{PhotoWeb spatial galaxy distribution}\label{sec:galaxy_spatial_distribution}

Given the narrow dispersion in the PhotoWeb redshift estimations we are able to assign accurate redshifts to \textit{individual galaxies}.  Figure \ref{fig:PhotoWeb_SDSS} shows the spatial distribution of galaxies with redshifts assigned using photoweb from the SDSS photo-$z$ catalogue computed by \citet{Csabai07}. We convert redshift to distance using the approximation in equation \ref{eq:red_dista}. There are $\sim2.2\times10^6$ galaxies with photo-$z$ estimates in this redshift range, compared to $\sim5\times10^5$ galaxies with spectroscopy. For comparison we also show the galaxy positions computed using the photo-$z$ and spectroscopic samples. There are no visible structures in the photo-$z$ galaxy distribution. In stark contrast the distribution of PhotoWeb-$z$ galaxies shows a rich cosmic web. There are roughly 5 times more galaxies in the PhotoWeb sample compared to the spectroscopic sample. While the redshift estimation with PhotoWeb is clearly superior there are also some visible artifacts. There is a lack of structures aligned with the LoS, reflecting the compression algorithm used to correct the FoG redshift distortions \citep{Jones10}. This artifact could be alleviated by computing the density field from the raw spectroscopic positions, but we would then have the risk of wrongly introducing galaxies inside voids. There are also practically no galaxies inside voids. This is the result of taking the largest peak in the PhotoWeb probability distribution as the solution instead of performing a montecarlo sampling over $P(z)_{\textrm{\tiny{web}}}$.  In the present work use direct assignment instead of a stochastic sampling for simplicity and due to our current lack of knowledge on the relation between the DTFE reconstructed galaxy density field and the un-biased matter distribution. 

Even when the density and geometry constraints give multiple narrow peaks in $P(z)_{\textrm{\tiny{web}}}$, this does not affect the shape of the reconstructed cosmic structures. This seems an advantage, but in reality this may introduce significant contamination for local environment studies. The intermittent nature of the peaks in the $P(z)_{\textrm{\tiny{web}}}$ of an individual galaxy is manifested as a degeneracy in the photo web redshift solution and this introduces a potentially unknown contamination by placing a galaxy inside a different environment to where it really belongs. Large errors in the PhotoWeb redshift estimation originate from galaxies that are wrongly assigned to a nearby cosmic structure instead of their correct position. In cases when the photo-$z$ $P(z)_{\textrm{\tiny{photo}}}$ has a large dispersion and there are dense structures far from the true galaxy's redshift, this error is particularly severe. On possible way to alleviate this is to include color information in the density estimation and in the assignment of the photo web redshift to individual galaxies. For instance, we expect red luminous galaxies to populate dense environments and blue star-forming to avoid them. This information can be used as a prior, at least in some cases, to further constrain the PhotoWeb PDF. This is, however, beyond the scope of the present work and will be explored in future papers.

%
\begin{figure*}
  \begin{center}
      \includegraphics[width=0.9\textwidth,angle=0.0]{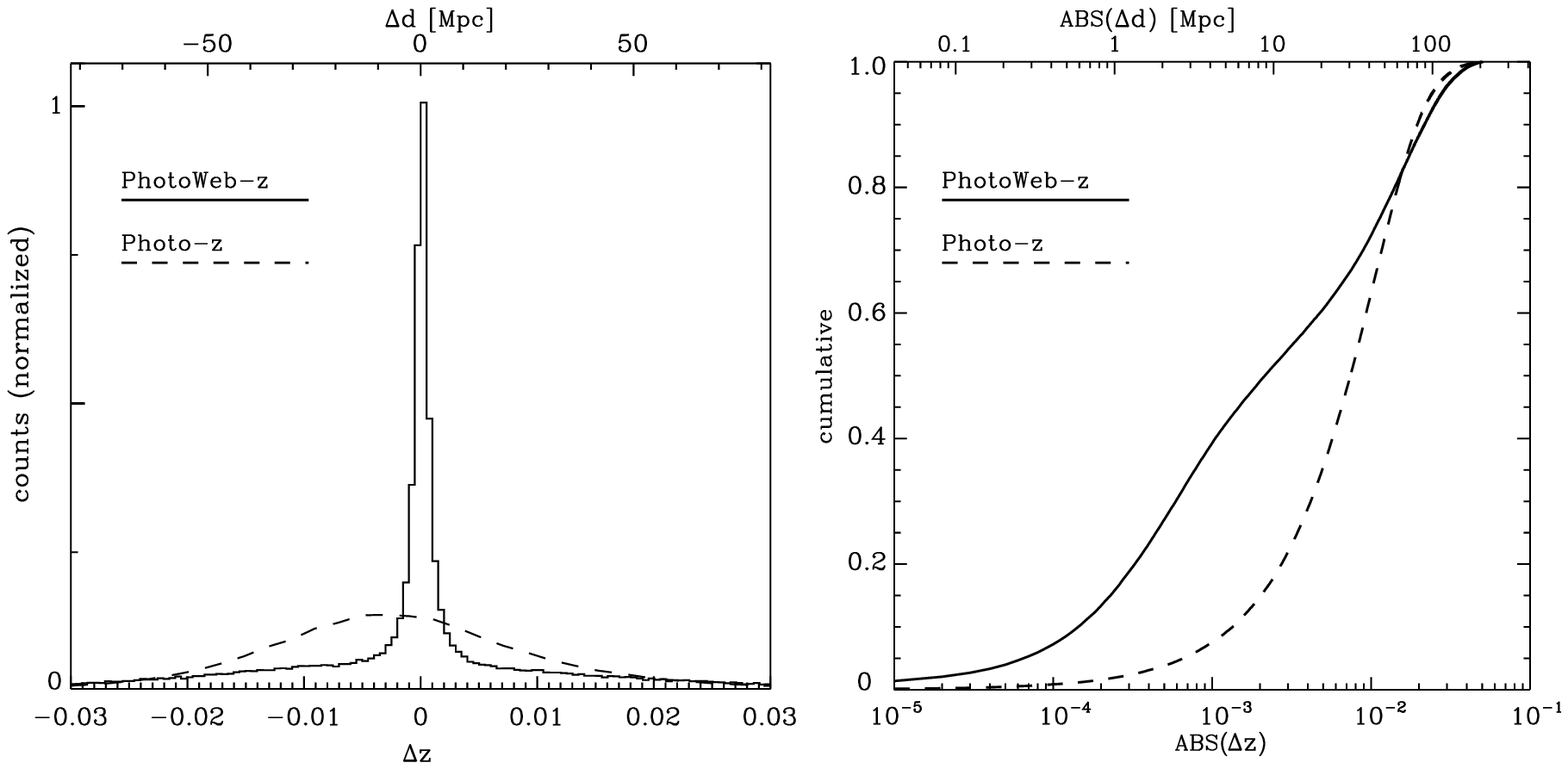}
  \end{center}
  \vspace{-0.5cm}
  \caption{\small {\bf Redshift estimation error.} Probability distribution (normalized) of the errors in PhotoWeb redshifts and photo-$z$ redshifts. The cumulative distribution (right panel) shows that almost half of the galaxies have PhotoWeb redshift estimates with errors one order of magnitude smaller than their corresponding photo-$z$.}
  \label{fig:photoweb_errors}
\end{figure*}

%
\subsubsection{Where is the information coming from?}

Figures \ref{fig:photoweb_diagram} and \ref{fig:PhotoWeb_SDSS} illustrate the fact that PhotoWeb does not add new information but it simply uses the information already present in the reference spectroscopic sample to guess the most likely locations of galaxies. PhotoWeb would not be able to reconstruct, say, a galaxy group for some (very unlikely) reason missed by the spectroscopic survey. However, the Delaunay tessellation performs a linear interpolation across structures which can be considered a first order reconstruction. Galaxy surveys like the SDSS provide a spectroscopic sampling dense enough to trace the small features of the cosmic web. Galaxies with no spectra are expected to fill the holes in between the sampling of galaxies with spectra.

%
\subsection{Estimated redshift error}\label{sec:errors}

Figure \ref{fig:photoweb_errors} shows the distribution of errors in the redshift (and distance) estimations from photo-$z$ and PhotoWeb. While the photo-$z$ errors have a broad Gaussian distribution, with a dispersion of $\Delta z \simeq 0.01-0.02$, equivalent to $d \simeq 40-80$ Mpc, the distribution of PhotoWeb errors has a narrow peak with dispersion $\Delta z \simeq 0.0006$, equivalent to $d \simeq 2.5$Mpc. The cumulative probability distribution shows that almost half of the galaxies have PhotoWeb redshift errors one order of magnitude smaller than photo-$z$. Compare this to typical photo-$z$ errors of $\Delta z \sim 0.02$ corresponding to a scale of $\sim 80$Mpc. The cumulative distribution shows a marginal excess of PhotoWeb compared to photo-$z$ for errors larger than $\sim 70$Mpc, which is of the order of the mean dispersion in the photo-$z$ errors. This corresponds to  $\sim \%15$ of the PhotoWeb galaxies having worse redshift estimations than photo-$z$. On the other side of the cumulative distribution we see that as the photo-$z$ errors decrease, their corresponding PhotoWeb errors also decrease at a higher rate. When the photo-$z$ errors are smaller than 10 Mpc the PhotoWeb errors are one order of magnitude smaller. Figure \ref{fig:photoweb_vs_photoz_errors} shows that for large photo-$z$ errors $\Delta z_{\textrm{\tiny{photo}}} > 0.01$ the PhotoWeb errors also perform poorly. Below $\Delta z_{\textrm{\tiny{photo}}}$ the PhotoWeb errors are smaller, almost independently of  $\Delta z_{\textrm{\tiny{photo}}}$ and with a peak around $\Delta z_{\textrm{\tiny{web}}} \sim 0.0007$.

%
\subsubsection{The nature of the PhotoWeb errors}\label{sec:error_model}

\begin{figure}
  \begin{center}
      \includegraphics[width=0.48\textwidth,angle=0.0]{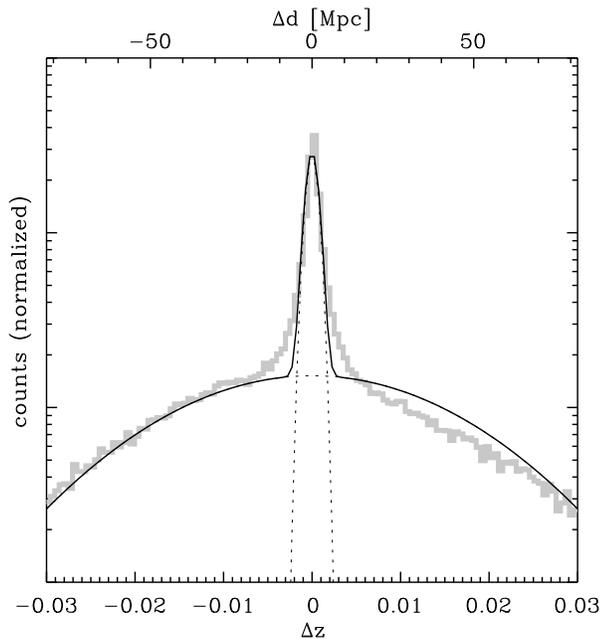}
  \end{center}
  \vspace{-0.5cm}
  \caption{\small {\bf PhotoWeb redshift error model.} The gray lines show the distribution of PhotoWeb errors. The black solid line shows the best fit from the two-component model in equation \ref{eq:two_component_model}. The narrow peak corresponds to small-scale errors from galaxies assigned to their correct cosmic structure. The broad peak corresponds to large-scale errors produced when PhotoWeb assigns a galaxy to a structure on the opposite side of the adjacent void (or even further) to the cosmic structure where the galaxy is truly located. The large-scale dispersion is close to the mean dispersion of the photo-$z$ errors.}
  \label{fig:photoweb_error_fit}
\end{figure}

The distribution of photo web errors, $\Delta z$ shown in Fig. \ref{fig:photoweb_errors} has a clearly non-gaussian shape. It consist of a high and narrow peak and tails extending well above what we expect from the exponential decline of the Gaussian function. The narrow and degenerate nature of the peaks in $P(z)_{\textrm{\tiny{web}}}$ results in either producing a very small redshift error when the galaxy is assigned to its correct cosmic structure, or a large redshift estimate when the galaxy is assigned to cosmic structures on the opposite sides of the adjacent voids to the structures where the galaxy is truly located. Following this rationale, we model the distribution of  photoweb errors into two components, one reflecting the small-scale errors from correctly assigned galaxies and another component describing the errors coming from assigning the galaxy to opposite cosmic structures

\begin{equation}
F(x) = F(x)_{\textrm{\tiny{s}}} + F(x)_{\textrm{\tiny{l}}} 
\end{equation}

\noindent where $x$ is the redshift error $\Delta z$ and $F(x)_{\textrm{\tiny{s}}}$,  $F(x)_{\textrm{\tiny{l}}}$ are the contributions from small and large scales respectively. Small-scale errors can be assumed to be Gaussian originating from uncertainties in the density estimation and non-linear processes such as peculiar velocities. Large-scale errors, on the other hand, originate from the discrete intermittent sampling of the Cosmic Web (see Fig. \ref{fig:photoweb_diagram} ).  An illuminating analogy is the intersection of a LoS with a Voronoi cellular distribution. The LoS will intersect the faces, edges and nodes of Voronoi cells \citep{Icke91}. The intersecting points will be separated by the empty spaces between intersected structures of the scale of the mean Voronoi cell size (\citet{Weygaert91,Weygaert94}, see also \citet{Weygaert02}). The LoS of target galaxies will intersect the cosmic structures at any position producing a roughly constant error distribution which is then weighted by the semi-Gaussian probability distribution of the photo-z estimation. For simplicity we assume a Gaussian distribution for both small-scale and large-scale errors. The total photoweb error is then modeled as:

%
\begin{figure}
  \begin{center}
      \includegraphics[width=0.48\textwidth,angle=0.0]{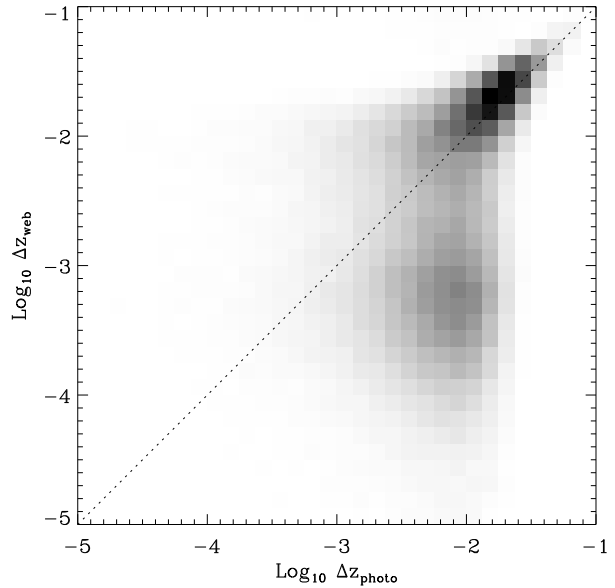}
  \end{center}
  \vspace{-0.5cm}
  \caption{\small {\bf PhotoWeb redshift error vs photo-$z$ redshift errors.} The gray scale indicates the counts inside each 2D bin. For large photo-$z$ errors ($\Delta z_{\textrm{\tiny{web}}} > 0.01$) the PhotoWeb errors also perform poorly. At smaller photo-$z$ errors ($\Delta z_{\textrm{\tiny{web}}} < 0.01$) the PhotoWeb errors are significantly smaller that photo-$z$.}
  \label{fig:photoweb_vs_photoz_errors}
\end{figure}

\begin{equation}\label{eq:two_component_model}
F(x) = A e^{-x^2 /2 \sigma_{\textrm{\tiny{s}}}^2} + B e^{-x^2 / 2 \sigma_{\textrm{\tiny{l}}}^2 } 
\end{equation}
\noindent The best fit to the distribution in Fig. \ref{fig:photoweb_error_fit} (normalized to 1) gives $A=1$, $ \sigma_{\textrm{\tiny{s}}} = 0.0007$, $B=0.08$ and $\sigma_{\textrm{\tiny{l}}}=0.0109593$. The small-scale dispersion in the redshift error corresponds to a physical scale of $\sim 2.5$Mpc, which is consistent to being originated from noise at the scale of cosmic structures such as walls, filaments and clusters, where most of the lines of sight intersect the cosmic web \citep{Aragon10c}. The large-scale redshift error dispersion is closer to the value of the mean redshift error in the photo-$z$ estimates used as input for photoweb $\sigma_{\textrm{\tiny{photo-z}}} \sim 0.02$. This two-component model is not strictly correct since the large-scale component $F(x)_{\textrm{\tiny{l}}}$ also contributes to small-scale errors (near the mean) from correctly assigned galaxies but nevertheless, it serves as a reference to compare the contribution of small and large-scale errors.

%

%
\begin{figure}
  \begin{center}
      \includegraphics[width=0.48\textwidth,angle=0.0]{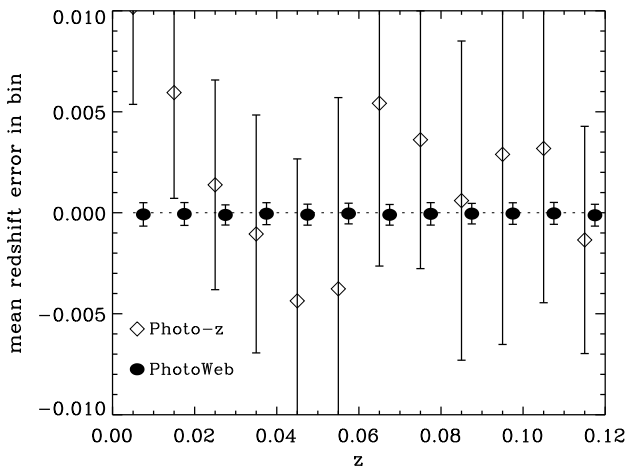}
  \end{center}
  \vspace{-0.5cm}
  \caption{\small {\bf Mean (truncated) redshift error.} The circles and diamonds mark the truncated mean redshift inside the corresponding bin and the error bars the dispersion inside the bin. The mean and dispersion where computed from their respective truncated (at $2\sigma$)  distribution. The mean redshifts from PhotoWeb are shifted to the right by 0.0025 in $z$ for clarity.}
  \label{fig:mean_redshift}
\end{figure}

%
\subsubsection{Mean redshift errors}\label{sec:mean_redshift_error}

An important quantity for cosmological studies is the mean redshift measured inside redshift bins. Even though photometric redshifts have large individual errors, their aggregate values approach the mean redshift inside a given redshift bin. Typical mean redshift errors from photo-$z$ are of the order of $0.001-0.01$ \citep{Carliles10,Rahman14}. However, the large dispersion in $P(z)_{\textrm{\tiny{photo}}}$ effectively acts like a smoothing and prevents us from resolving features in the redshift distribution at scales smaller than the typical redshift uncertainties, which for the SDSS photo-$z$ sample is $\sim 0.01-0.02$.

Figure \ref{fig:mean_redshift} shows the truncated mean of the redshift errors computed in several redshift bins. PhotoWeb is sensitive to biased photo-$z$ estimates such as the SDSS sample we used where there are systematically higher estimations at low redshift as shown in Fig. \ref{fig:photoweb_vs_spec_2d_histogram}. In order to correct for this effect we ignore errors larger than two standard deviations of the respective error distribution (see Sec. \ref{sec:errors} and Fig. \ref{fig:photoweb_errors}). The mean truncated PhotoWeb redshift errors are indistinguishable  from the spectroscopic redshifts with values in the range $\Delta z \simeq 10^{-5}-10^{-4}$. These values are more than one order of magnitude smaller compared to photo-$z$ using only color information (our reference sample) and even photo-$z$ estimates reported in the literature after including galaxy correlation information $\Delta z = 0.001-0.01$ \citep{Rahman14}. Note, however that the mean errors were computed from the truncated distribution, this assumes that we have access to unbiased photo-$z$ estimates which, given the recent advances in photo-$z$ techniques is feasible.


\section{Conclusions and future work}

The method presented here, PhotoWeb,  offers an unparalleled accuracy in redshift estimation by combining the broad unique peak in the probability distribution obtained from photo-$z$ estimations with the narrow but degenerated solutions given by the cosmic structures in the galaxy distribution. The current implementation of PhotoWeb demonstrates the constraining power of the cosmic web on galaxy positions along the LoS, achieving redshift errors of the order to $\Delta z \sim 0.0007$ and mean redshifts of the order of $\sim 10^{-5}$. 

By using PhotoWeb we can increase the number of galaxies with accurate redshifts in SDSS from $\sim 10^6$ in the spectroscopic sample to $\sim 5\times10^6$ (including galaxies with small photo-$z$ uncertainty). This factor of 5 can be further increased as better photo-$z$ estimates become available. This unprecedented galaxy sampling will allow environmental studies of galaxy formation and evolution previously unfeasible with either spectroscopic or photometric redshifts.

The major limitation of PhotoWeb (in its current implementation) is the fraction of catastrophic errors occurring when a galaxy is assigned to a different cosmic structure where it belongs. We note that this fraction is essentially equal to the one derived from photo-$z$. When PhotoWeb succeeds in identifying the galaxy's host structure it does so superbly. When it fails it does so in a similar degree as photo-$z$. This is in part due to the fact that when photo-$z$ fails catastrophically there is no way PhotoWeb can identify the correct cosmic structure. The error in the peak of $P(z)_{\textrm{\tiny{photo}}}$ as well as its dispersion must be of the order of the typical void size for PhotoWeb to succeed. A major challenge is then the reduction of the fraction of galaxies  erroneously assigned to nearby cosmic structures. This may seem an impossible task without the use of better photometric redshifts but there is in fact more information available that we have not yet exploited in PhotoWeb. In particular galaxy color and morphology, used as a prior, and can play a significant role in further constraining $P(z)_{\textrm{\tiny{photo}}}$. Galaxies populate the cosmic web following well known rules such as the color and morphology-density relation \citep{Dressler80}. Our lack of knowledge on the underlying physical process does not prevent the use of these empirical relations to constrain galaxy locations along the LoS.

%
\subsection{Future improvements}

In the present implementation of PhotoWeb we use a spectroscopic sample to derive $P(z)_{\textrm{\tiny{den}}}$ and $P(z)_{\textrm{\tiny{geo}}}$. However, for surveys with a large number of narrow-band filters, where the photo-$z$ estimates have an uncertainty of less than a few megaparsecs it may be possible to use the photo-$z$ sample as the reference and apply PhotoWeb iteratively. This is possible because walls, filaments and clusters are located at the boundaries of voids, and voids are relatively easy to identify as local minima in the density field.

The density field used in this work was computed in a regular cubic grid. A polar grid is an alternative option which has some desirable features like a natural degrading of spatial resolution with increasing distance, providing some compensation for the galaxy selection function and a better fit to the survey geometry. We will explore this possibility in a forthcoming paper.

The current implementation of PhotoWeb only takes advantage of the spatial coherence of the cosmic web. However, galaxies also present clustering based on their color and morphology. The distribution of galaxies has a well-known color segregation, red early type galaxies tend to populate high density regions while blue late-type avoid them and are mostly found in the intermediate and low density regions \citep{Dressler80}. We can use this information to further constrain the location of a galaxy on the LoS. For instance if the target galaxy is classified as late-type and is blue ($u-r < 2.2$) then high-density regions in the LoS density field will be assigned a low probability. The simplest way to do this would be to separate galaxies in color bins and compute density fields for each color bin separately but this would significantly reduce the number of available galaxies to reconstruct the density field. Instead we can weight each galaxy according to their color and compute two separate density fields with \textit{enhanced} red/blue galaxy populations as:

\begin{eqnarray}
\rho_{\textrm{\tiny{red}}}  & = & (u-r) \; \rho, \;\;\;\;\;\; \textrm{if} \;\;\;u-r > 2.2\\
\rho_{\textrm{\tiny{blue}}} & = & (r-u) \; \rho, \;\;\;\;\;\; \textrm{if} \;\;\;u-r < 2.2
\end{eqnarray}

\noindent Where $\rho_{\textrm{\tiny{red}}}, \rho_{\textrm{\tiny{blue}}}$ are the color-weighted density estimations and $\rho$ is the original density estimate at the position of the galaxy computed with the DTFE. 


Photometric redshift estimation methods are all based on a mapping between the observed properties of galaxies and some known model. The mapping is usually straightforward and applicable to the galaxy sample. However, as early examples of redshift estimation have shown, it is possible, at least in some cases, to use extra information to further constrain the location of a galaxy, for instance by looking at signs of interaction with a companion of known redshift, or by comparing the color/morphology of the galaxy with the galaxies along its LoS to, say, discard massive luminous elliptical galaxies from the centers of voids. These are, so far, anecdotical cases but nothing prevents us from envisioning a general framework, extending the notion introduced by \citet{Budavari09}, in which every possible piece of information is used to estimate accurate redshifts. We currently do not have the technology to do this in an automated way, however, new advances in image processing, machine learning and automated pattern recognition can enable this kind of analysis in the near future.


\section{Acknowledgements}
This research was partially funded by a UC Riverside Big Data seed grant as part of the Multidisciplinary Image Processing Laboratory (MIPLab).  M.A. Aragon-Calvo would like to thank Mark Neyrinck for useful comments on the early manuscript.

Funding for the SDSS and SDSS-II has been provided by the Alfred P. Sloan Foundation, the Participating Institutions, the National Science Foundation, the U.S. Department of Energy, the National Aeronautics and Space Administration, the Japanese Monbukagakusho, the Max Planck Society, and the Higher Education Funding Council for England. The SDSS Web Site is http://www.sdss.org/.

The SDSS is managed by the Astrophysical Research Consortium for the Participating Institutions. The Participating Institutions are the American Museum of Natural History, Astrophysical Institute Potsdam, University of Basel, University of Cambridge, Case Western Reserve University, University of Chicago, Drexel University, Fermilab, the Institute for Advanced Study, the Japan Participation Group, Johns Hopkins University, the Joint Institute for Nuclear Astrophysics, the Kavli Institute for Particle Astrophysics and Cosmology, the Korean Scientist Group, the Chinese Academy of Sciences (LAMOST), Los Alamos National Laboratory, the Max-Planck-Institute for Astronomy (MPIA), the Max- Planck-Institute for Astrophysics (MPA), New Mexico State University, Ohio State University, University of Pittsburgh, University of Portsmouth, Princeton University, the United States Naval Observatory, and the University of Washington.

\bibliography{refs} 

\begin{thebibliography}{56}
\expandafter\ifx\csname natexlab\endcsname\relax\def\natexlab#1{#1}\fi

\bibitem[{{Abazajian} {et~al}\mbox{.}(2009){Abazajian}, {Adelman-McCarthy},
  {Ag{\"u}eros}, {Allam}, {Allende Prieto}, {An}, {Anderson}, {Anderson},
  {Annis}, {Bahcall}, \& et~al.}]{Abazajian09}
{Abazajian} K.~N. {et~al.}, 2009, \apjs, 182, 543

\bibitem[{{Aragon-Calvo}(2012)}]{Aragon13b}
{Aragon-Calvo} M.~A., 2012, ArXiv e-prints

\bibitem[{{Arag{\'o}n-Calvo} {et~al}\mbox{.}(2010){Arag{\'o}n-Calvo}, {Platen},
  {van de Weygaert}, \& {Szalay}}]{Aragon10}
{Arag{\'o}n-Calvo} M.~A., {Platen} E., {van de Weygaert} R., {Szalay} A.~S.,
  2010, \apj, 723, 364

\bibitem[{{Aragon-Calvo} \& {Szalay}(2013)}]{Aragon13a}
{Aragon-Calvo} M.~A., {Szalay} A.~S., 2013, \mnras, 428, 3409

\bibitem[{{Arag{\'o}n-Calvo}, {van de Weygaert} \&
  {Jones}(2010){Arag{\'o}n-Calvo}, {van de Weygaert}, \& {Jones}}]{Aragon10c}
{Arag{\'o}n-Calvo} M.~A., {van de Weygaert} R., {Jones} B.~J.~T., 2010, \mnras,
  408, 2163

\bibitem[{{Ben{\'{\i}}tez}(2000)}]{Benitez00}
{Ben{\'{\i}}tez} N., 2000, \apj, 536, 571

\bibitem[{{Berlind} {et~al}\mbox{.}(2006){Berlind}, {Frieman}, {Weinberg},
  {Blanton}, {Warren}, {Abazajian}, {Scranton}, {Hogg}, {Scoccimarro},
  {Bahcall}, {Brinkmann}, {Gott}, {Kleinman}, {Krzesinski}, {Lee}, {Miller},
  {Nitta}, {Schneider}, {Tucker}, {Zehavi}, \& {SDSS
  Collaboration}}]{Berlind06}
{Berlind} A.~A. {et~al.}, 2006, \apjs, 167, 1

\bibitem[{Beucher \& Lantuejoul(1979)}]{Beucher79}
Beucher S., Lantuejoul C., 1979, in International Workshop on Image Processing:
  Real-time Edge and Motion Detection/Estimation, Rennes, France.

\bibitem[{{Bond}, {Kofman} \& {Pogosyan}(1996){Bond}, {Kofman}, \&
  {Pogosyan}}]{Bond96}
{Bond} J.~R., {Kofman} L., {Pogosyan} D., 1996, \nat, 380, 603

\bibitem[{{Broadhurst} {et~al}\mbox{.}(1990){Broadhurst}, {Ellis}, {Koo}, \&
  {Szalay}}]{Broadhurst90}
{Broadhurst} T.~J., {Ellis} R.~S., {Koo} D.~C., {Szalay} A.~S., 1990, \nat,
  343, 726

\bibitem[{{Budav{\'a}ri}(2009)}]{Budavari09}
{Budav{\'a}ri} T., 2009, \apj, 695, 747

\bibitem[{{Carliles} {et~al}\mbox{.}(2010){Carliles}, {Budav{\'a}ri}, {Heinis},
  {Priebe}, \& {Szalay}}]{Carliles10}
{Carliles} S., {Budav{\'a}ri} T., {Heinis} S., {Priebe} C., {Szalay} A.~S.,
  2010, \apj, 712, 511

\bibitem[{{Carron}(2011)}]{Carron11}
{Carron} J., 2011, \apj, 738, 86

\bibitem[{{Carron} \& {Neyrinck}(2012)}]{Carron12}
{Carron} J., {Neyrinck} M.~C., 2012, \apj, 750, 28

\bibitem[{{Cautun} {et~al}\mbox{.}(2014){Cautun}, {van de Weygaert}, {Jones},
  \& {Frenk}}]{Cautun14}
{Cautun} M., {van de Weygaert} R., {Jones} B.~J.~T., {Frenk} C.~S., 2014,
  \mnras, 441, 2923

\bibitem[{{Coles} \& {Jones}(1991)}]{Coles91}
{Coles} P., {Jones} B., 1991, \mnras, 248, 1

\bibitem[{{Connolly} {et~al}\mbox{.}(1995){Connolly}, {Csabai}, {Szalay},
  {Koo}, {Kron}, \& {Munn}}]{Connolly95}
{Connolly} A.~J., {Csabai} I., {Szalay} A.~S., {Koo} D.~C., {Kron} R.~G.,
  {Munn} J.~A., 1995, \aj, 110, 2655

\bibitem[{{Csabai} {et~al}\mbox{.}(2007){Csabai}, {Dobos}, {Trencs{\'e}ni},
  {Herczegh}, {J{\'o}zsa}, {Purger}, {Budav{\'a}ri}, \& {Szalay}}]{Csabai07}
{Csabai} I., {Dobos} L., {Trencs{\'e}ni} M., {Herczegh} G., {J{\'o}zsa} P.,
  {Purger} N., {Budav{\'a}ri} T., {Szalay} A.~S., 2007, Astronomische
  Nachrichten, 328, 852

\bibitem[{{Cunha} {et~al}\mbox{.}(2009){Cunha}, {Lima}, {Oyaizu}, {Frieman}, \&
  {Lin}}]{Cunha09}
{Cunha} C.~E., {Lima} M., {Oyaizu} H., {Frieman} J., {Lin} H., 2009, \mnras,
  396, 2379

\bibitem[{{Dressler}(1980)}]{Dressler80}
{Dressler} A., 1980, \apj, 236, 351

\bibitem[{{Efstathiou} \& {Moody}(2001)}]{Efstathiou01}
{Efstathiou} G., {Moody} S.~J., 2001, \mnras, 325, 1603

\bibitem[{{Fry}(1984)}]{Fry84}
{Fry} J.~N., 1984, \apjl, 277, L5

\bibitem[{{Goldberg} \& {Vogeley}(2004)}]{Goldberg04}
{Goldberg} D.~M., {Vogeley} M.~S., 2004, \apj, 605, 1

\bibitem[{{Gottl{\"o}ber} {et~al}\mbox{.}(2003){Gottl{\"o}ber}, {{\L}okas},
  {Klypin}, \& {Hoffman}}]{Gottlober03}
{Gottl{\"o}ber} S., {{\L}okas} E.~L., {Klypin} A., {Hoffman} Y., 2003, \mnras,
  344, 715

\bibitem[{{Hildebrandt} {et~al}\mbox{.}(2010){Hildebrandt}, {Arnouts}, {Capak},
  {Moustakas}, {Wolf}, {Abdalla}, {Assef}, {Banerji}, {Ben{\'{\i}}tez},
  {Brammer}, {Budav{\'a}ri}, {Carliles}, {Coe}, {Dahlen}, {Feldmann}, {Gerdes},
  {Gillis}, {Ilbert}, {Kotulla}, {Lahav}, {Li}, {Miralles}, {Purger},
  {Schmidt}, \& {Singal}}]{Hildebrandt10}
{Hildebrandt} H. {et~al.}, 2010, \aap, 523, A31

\bibitem[{{Huchra} \& {Geller}(1982)}]{Huchra82}
{Huchra} J.~P., {Geller} M.~J., 1982, \apj, 257, 423

\bibitem[{{Icke}(1984)}]{Icke84}
{Icke} V., 1984, \mnras, 206, 1P

\bibitem[{{Icke} \& {van de Weygaert}(1991)}]{Icke91}
{Icke} V., {van de Weygaert} R., 1991, \qjras, 32, 85

\bibitem[{{Jasche} \& {Wandelt}(2012)}]{Jasche12}
{Jasche} J., {Wandelt} B.~D., 2012, \mnras, 425, 1042

\bibitem[{{Jones}, {van de Weygaert} \& {Arag{\'o}n-Calvo}(2010){Jones}, {van
  de Weygaert}, \& {Arag{\'o}n-Calvo}}]{Jones10}
{Jones} B.~J.~T., {van de Weygaert} R., {Arag{\'o}n-Calvo} M.~A., 2010, \mnras,
  408, 897

\bibitem[{{Kaiser}(1984)}]{Kaiser84}
{Kaiser} N., 1984, \apjl, 284, L9

\bibitem[{{Kaiser}(1987)}]{Kaiser87}
{Kaiser} N., 1987, \mnras, 227, 1

\bibitem[{{Kirshner} {et~al}\mbox{.}(1981){Kirshner}, {Oemler}, {Schechter}, \&
  {Shectman}}]{Kirshner81}
{Kirshner} R.~P., {Oemler}, Jr. A., {Schechter} P.~L., {Shectman} S.~A., 1981,
  \apjl, 248, L57

\bibitem[{{Kova{\v c}} {et~al}\mbox{.}(2010){Kova{\v c}}, {Lilly}, {Cucciati},
  {Porciani}, {Iovino}, {Zamorani}, {Oesch}, {Bolzonella}, {Knobel},
  {Finoguenov}, {Peng}, {Carollo}, {Pozzetti}, {Caputi}, {Silverman}, {Tasca},
  {Scodeggio}, {Vergani}, {Scoville}, {Capak}, {Contini}, {Kneib}, {Le
  F{\`e}vre}, {Mainieri}, {Renzini}, {Bardelli}, {Bongiorno}, {Coppa}, {de la
  Torre}, {de Ravel}, {Franzetti}, {Garilli}, {Guzzo}, {Kampczyk},
  {Lamareille}, {Le Borgne}, {Le Brun}, {Maier}, {Mignoli}, {Pello}, {Perez
  Montero}, {Ricciardelli}, {Tanaka}, {Tresse}, {Zucca}, {Abbas}, {Bottini},
  {Cappi}, {Cassata}, {Cimatti}, {Fumana}, {Koekemoer}, {Maccagni}, {Marinoni},
  {McCracken}, {Memeo}, {Meneux}, \& {Scaramella}}]{Kovac10}
{Kova{\v c}} K. {et~al.}, 2010, \apj, 708, 505

\bibitem[{{M{\'e}nard} {et~al}\mbox{.}(2013){M{\'e}nard}, {Scranton},
  {Schmidt}, {Morrison}, {Jeong}, {Budavari}, \& {Rahman}}]{Menard13}
{M{\'e}nard} B., {Scranton} R., {Schmidt} S., {Morrison} C., {Jeong} D.,
  {Budavari} T., {Rahman} M., 2013, ArXiv e-prints

\bibitem[{{Mo} \& {White}(1996)}]{Mo96}
{Mo} H.~J., {White} S.~D.~M., 1996, \mnras, 282, 347

\bibitem[{{Mobasher} {et~al}\mbox{.}(2007){Mobasher}, {Capak}, {Scoville},
  {Dahlen}, {Salvato}, {Aussel}, {Thompson}, {Feldmann}, {Tasca}, {Le Fevre},
  {Lilly}, {Carollo}, {Kartaltepe}, {McCracken}, {Mould}, {Renzini}, {Sanders},
  {Shopbell}, {Taniguchi}, {Ajiki}, {Shioya}, {Contini}, {Giavalisco},
  {Ilbert}, {Iovino}, {Le Brun}, {Mainieri}, {Mignoli}, \&
  {Scodeggio}}]{Mobasher07}
{Mobasher} B. {et~al.}, 2007, \apjs, 172, 117

\bibitem[{{Newman}(2008)}]{Newman08}
{Newman} J.~A., 2008, \apj, 684, 88

\bibitem[{{Neyrinck} {et~al}\mbox{.}(2014){Neyrinck}, {Arag{\'o}n-Calvo},
  {Jeong}, \& {Wang}}]{Neyrinck14}
{Neyrinck} M.~C., {Arag{\'o}n-Calvo} M.~A., {Jeong} D., {Wang} X., 2014,
  \mnras, 441, 646

\bibitem[{{Neyrinck}, {Szapudi} \& {Szalay}(2009){Neyrinck}, {Szapudi}, \&
  {Szalay}}]{Neyrinck09}
{Neyrinck} M.~C., {Szapudi} I., {Szalay} A.~S., 2009, \apjl, 698, L90

\bibitem[{Nguyen(2007)}]{Nguyen07}
Nguyen H., 2007, GPU Gems 3, 1st edn. Addison-Wesley Professional

\bibitem[{{Platen}, {van de Weygaert} \& {Jones}(2007){Platen}, {van de
  Weygaert}, \& {Jones}}]{Platen07}
{Platen} E., {van de Weygaert} R., {Jones} B.~J.~T., 2007, \mnras, 380, 551

\bibitem[{{Platen}, {van de Weygaert} \& {Jones}(2008){Platen}, {van de
  Weygaert}, \& {Jones}}]{Platen08}
{Platen} E., {van de Weygaert} R., {Jones} B.~J.~T., 2008, \mnras, 387, 128

\bibitem[{{Platen} {et~al}\mbox{.}(2011){Platen}, {van de Weygaert}, {Jones},
  {Vegter}, \& {Calvo}}]{Platen11}
{Platen} E., {van de Weygaert} R., {Jones} B.~J.~T., {Vegter} G., {Calvo}
  M.~A.~A., 2011, \mnras, 416, 2494

\bibitem[{{Rahman} {et~al}\mbox{.}(2014){Rahman}, {M{\'e}nard}, {Scranton},
  {Schmidt}, \& {Morrison}}]{Rahman14}
{Rahman} M., {M{\'e}nard} B., {Scranton} R., {Schmidt} S.~J., {Morrison} C.~B.,
  2014, ArXiv e-prints

\bibitem[{{Schaap} \& {van de Weygaert}(2000)}]{Schaap00}
{Schaap} W.~E., {van de Weygaert} R., 2000, \aap, 363, L29

\bibitem[{{Schneider} {et~al}\mbox{.}(2006){Schneider}, {Knox}, {Zhan}, \&
  {Connolly}}]{Schneider06}
{Schneider} M., {Knox} L., {Zhan} H., {Connolly} A., 2006, \apj, 651, 14

\bibitem[{{Steinberg} {et~al}\mbox{.}(2014){Steinberg}, {Yalinewich}, {Sari},
  \& {Duffell}}]{Steinberg14}
{Steinberg} E., {Yalinewich} A., {Sari} R., {Duffell} P., 2014, ArXiv e-prints

\bibitem[{{Szapudi}, {Szalay} \& {Boschan}(1991){Szapudi}, {Szalay}, \&
  {Boschan}}]{Szapudi91}
{Szapudi} I., {Szalay} A.~S., {Boschan} P., 1991, in Traces of the Primordial
  Structure in the Universe, pp. 39--56

\bibitem[{{Tegmark} {et~al}\mbox{.}(2004){Tegmark}, {Blanton}, {Strauss},
  {Hoyle}, {Schlegel}, {Scoccimarro}, {Vogeley}, {Weinberg}, {Zehavi},
  {Berlind}, {Budavari}, {Connolly}, {Eisenstein}, {Finkbeiner}, {Frieman},
  {Gunn}, {Hamilton}, {Hui}, {Jain}, {Johnston}, {Kent}, {Lin}, {Nakajima},
  {Nichol}, {Ostriker}, {Pope}, {Scranton}, {Seljak}, {Sheth}, {Stebbins},
  {Szalay}, {Szapudi}, {Verde}, {Xu}, {Annis}, {Bahcall}, {Brinkmann},
  {Burles}, {Castander}, {Csabai}, {Loveday}, {Doi}, {Fukugita}, {Gott},
  {Hennessy}, {Hogg}, {Ivezi{\'c}}, {Knapp}, {Lamb}, {Lee}, {Lupton}, {McKay},
  {Kunszt}, {Munn}, {O'Connell}, {Peoples}, {Pier}, {Richmond}, {Rockosi},
  {Schneider}, {Stoughton}, {Tucker}, {Vanden Berk}, {Yanny}, {York}, \& {SDSS
  Collaboration}}]{Tegmark04}
{Tegmark} M. {et~al.}, 2004, \apj, 606, 702

\bibitem[{{van de Weygaert}(1991)}]{Weygaert91}
{van de Weygaert} R., 1991, \mnras, 249, 159

\bibitem[{{van de Weygaert}(1994)}]{Weygaert94}
{van de Weygaert} R., 1994, \aap, 283, 361

\bibitem[{{van de Weygaert}(2002)}]{Weygaert02}
{van de Weygaert} R., 2002, ArXiv Astrophysics e-prints

\bibitem[{{York} {et~al}\mbox{.}(2000){York}, {Adelman}, {Anderson},
  {Anderson}, {Annis}, {Bahcall}, {Bakken}, {Barkhouser}, {Bastian}, {Berman},
  {Boroski}, {Bracker}, {Briegel}, {Briggs}, {Brinkmann}, {Brunner}, {Burles},
  {Carey}, {Carr}, {Castander}, {Chen}, {Colestock}, {Connolly}, {Crocker},
  {Csabai}, {Czarapata}, {Davis}, {Doi}, {Dombeck}, {Eisenstein}, {Ellman},
  {Elms}, {Evans}, {Fan}, {Federwitz}, {Fiscelli}, {Friedman}, {Frieman},
  {Fukugita}, {Gillespie}, {Gunn}, {Gurbani}, {de Haas}, {Haldeman}, {Harris},
  {Hayes}, {Heckman}, {Hennessy}, {Hindsley}, {Holm}, {Holmgren}, {Huang},
  {Hull}, {Husby}, {Ichikawa}, {Ichikawa}, {Ivezi{\'c}}, {Kent}, {Kim},
  {Kinney}, {Klaene}, {Kleinman}, {Kleinman}, {Knapp}, {Korienek}, {Kron},
  {Kunszt}, {Lamb}, {Lee}, {Leger}, {Limmongkol}, {Lindenmeyer}, {Long},
  {Loomis}, {Loveday}, {Lucinio}, {Lupton}, {MacKinnon}, {Mannery}, {Mantsch},
  {Margon}, {McGehee}, {McKay}, {Meiksin}, {Merelli}, {Monet}, {Munn},
  {Narayanan}, {Nash}, {Neilsen}, {Neswold}, {Newberg}, {Nichol}, {Nicinski},
  {Nonino}, {Okada}, {Okamura}, {Ostriker}, {Owen}, {Pauls}, {Peoples},
  {Peterson}, {Petravick}, {Pier}, {Pope}, {Pordes}, {Prosapio},
  {Rechenmacher}, {Quinn}, {Richards}, {Richmond}, {Rivetta}, {Rockosi},
  {Ruthmansdorfer}, {Sandford}, {Schlegel}, {Schneider}, {Sekiguchi}, {Sergey},
  {Shimasaku}, {Siegmund}, {Smee}, {Smith}, {Snedden}, {Stone}, {Stoughton},
  {Strauss}, {Stubbs}, {SubbaRao}, {Szalay}, {Szapudi}, {Szokoly}, {Thakar},
  {Tremonti}, {Tucker}, {Uomoto}, {Vanden Berk}, {Vogeley}, {Waddell}, {Wang},
  {Watanabe}, {Weinberg}, {Yanny}, {Yasuda}, \& {SDSS Collaboration}}]{York00}
{York} D.~G. {et~al.}, 2000, \aj, 120, 1579

\bibitem[{{Zeldovich}, {Einasto} \& {Shandarin}(1982){Zeldovich}, {Einasto}, \&
  {Shandarin}}]{Zeldovich82}
{Zeldovich} I.~B., {Einasto} J., {Shandarin} S.~F., 1982, \nat, 300, 407

\bibitem[{{Zel'dovich}(1970)}]{Zeldovich70}
{Zel'dovich} Y.~B., 1970, \aap, 5, 84

\end{thebibliography}
\bibliographystyle{mn2e}   

\appendix

\section{Delaunay-based structure reconstruction}\label{app:delaunay_reconstruction}

%
\begin{figure*}
  \begin{center}
       \includegraphics[width=0.65\textwidth,angle=0.0]{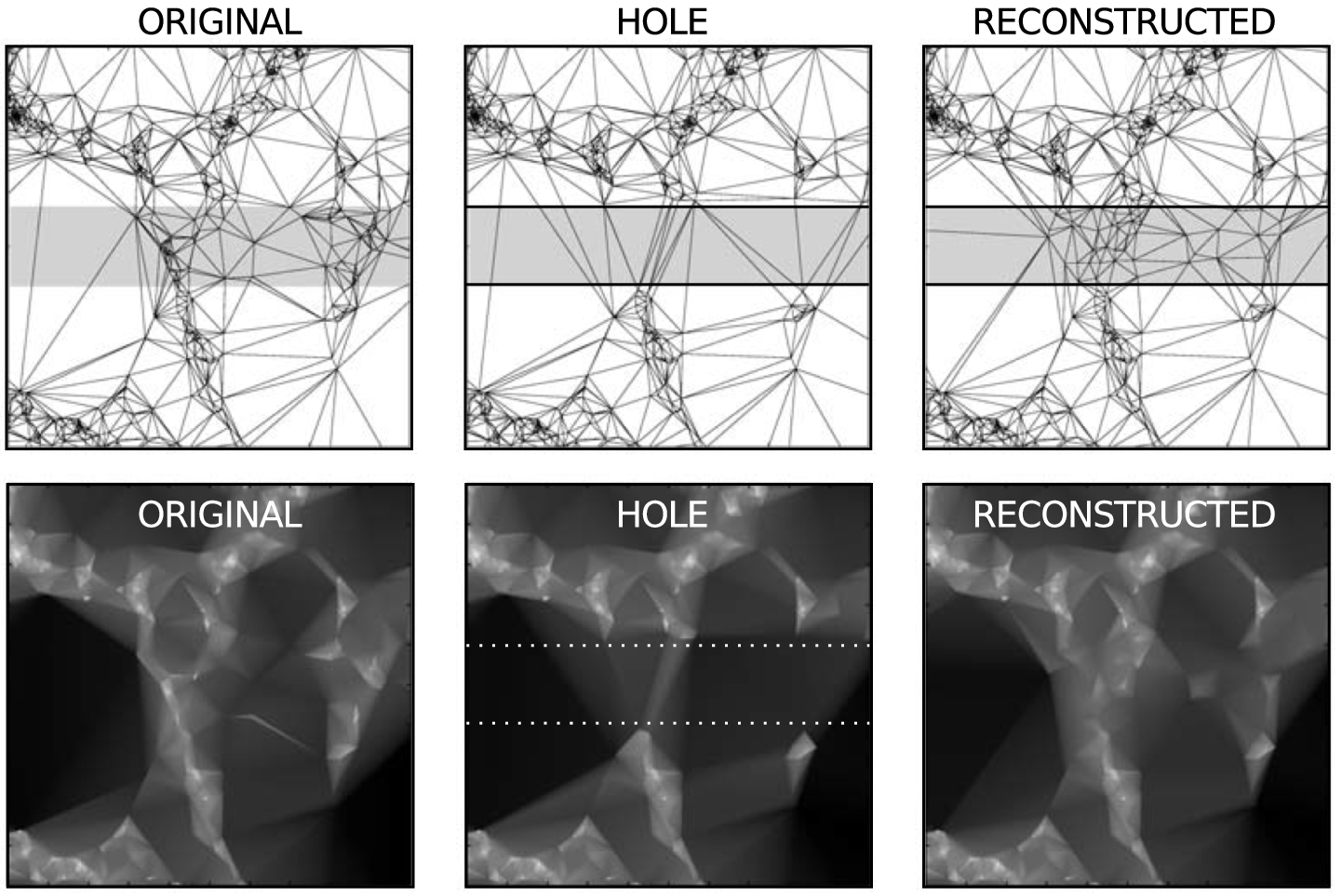}
  \end{center}
  \vspace{-0.5cm}
  \caption{\small {\bf Delaunay-based structure reconstruction.} The left panels show the Delaunay tessellation (top) and reconstructed density field (bottom) corresponding to the biased particle distribution in an N-body simulation. Central panels show the corresponding Delaunay tessellation and density field after a hole has been cut from the original particle distribution. The hole is indicated by the horizontal shaded area crossing the top panels. By sampling the density field inside the hole with new particles matching the mean number density outside the hole we are able to reconstruct to a remarkable degree the original density field.}
  \label{fig:DTFI}
\end{figure*}

The Delaunay tessellation not only is able to compute density estimates and interpolate their value at any point but it can also interpolate anisotropic features in the galaxy distribution. Figure \ref{fig:DTFI} shows the reconstruction of the (2D) density field from a particle distribution before and after particles have been removed to form a hole. The Delaunay tessellation interpolates the space in-between the hole joining and reconstructing the filament previously truncated by the hole. We exploit this property of the Delaunay tessellation to interpolate the space between galaxies in sparsely sampled spectroscopic surveys and reconstruct the underlying density field. It is important to note that the Delaunay tessellation interpolates the density field in a linear fashion. This limits its application to linear features. However, as long as the spectroscopic sampling used to construct the tessellation can sample the features we want to reconstruct, i.e. if its mean inter-galaxy separation is smaller than the typical scale of the cosmic structures, then the linear interpolation is a good approximation.

\section{SQL query}\label{app:sql}

The spectroscopic and photometric redshifts were obtained using the SDSS CASJOBS service with the following query:

\begin{verbatim}
SELECT s.ra, s.dec,s.z, p.z,p.zErr,
     p.absMagI,p.absMagR,p.absMagG,p.absMagU
FROM Photoz AS p
JOIN SpecPhoto as s
ON   p.ObjID = s.ObjID
WHERE
     p.z BETWEEN 0 and 0.15
     AND p.nnIsInside = 1 
     AND p.nnCount    > 95 
     AND p.zErr BETWEEN 0 AND 0.02
\end{verbatim}

\noindent From this master catalogue we generated subsamples as needed. The photometric redshift catalogue from \citet{Csabai07} was queried in the same way as above but without performing the JOIN with the spectroscopic sample. The spectroscopic galaxy catalogue used to generate density fields was produced using the following query:

\begin{verbatim}
SELECT  p.ObjID, p.ra, p.dec, s.z, s.zErr, s.zConf, 
     s.cx, s.cy, s.cz,
     p.dered_u, p.dered_g, p.dered_r, ...
     p.isoPhi_u, p.isoPhi_g, p.isoPhi_r, ...
     p.isoA_u, p.isoB_u, p.isoA_g, p.isoB_g, ...
     p.petroR50_u, p.petroR90_u, p.petroR50_g, ...
     p.petroMag_u,p.petroMag_g,p.petroMag_r,...
     p.petroRad_u,p.petroRad_g,p.petroRad_r,...
     s.specclass,
     s.eClass
INTO myDB.galaxies_dr9
FROM SpecObj  AS s, 
     PhotoObj AS p
WHERE s.SpecObjID = p.SpecObjID
 \end{verbatim}

\section{Selection function weighting}\label{app:selection_function}

%
\begin{figure}
  \begin{center}
       \includegraphics[width=0.48\textwidth,angle=0.0]{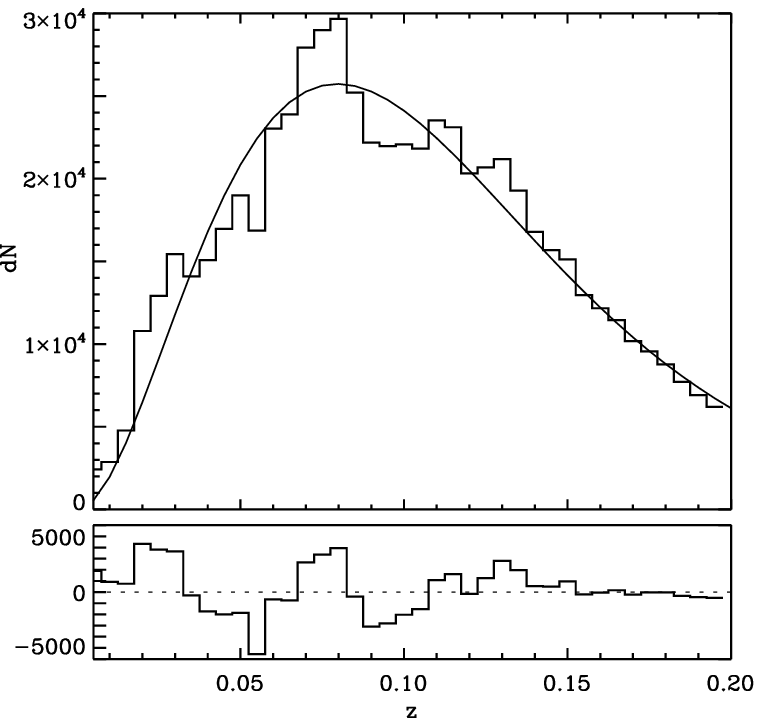}
  \end{center}
  \vspace{-0.5cm}
  \caption{Distribution of galaxies with redshift.}
  \label{fig:SDSS_selection_function}
\end{figure}

%
\begin{figure}
  \begin{center}
       \includegraphics[width=0.4\textwidth,angle=0.0]{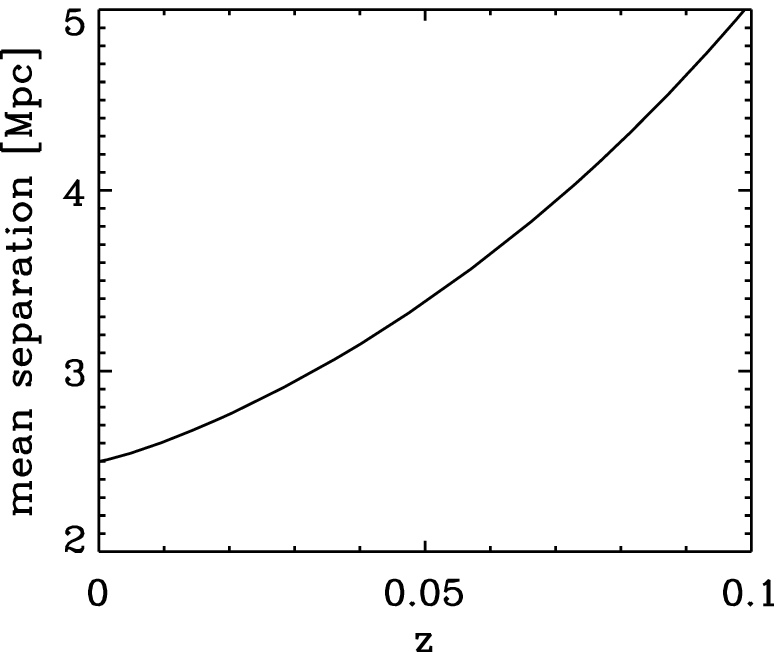}
  \end{center}
  \vspace{-0.5cm}
  \caption{Mean inter-galaxy separation as function of redshift.}
  \label{fig:SDSS_mean_separation}
\end{figure}

In order to take full advantage of all the galaxies in the sample, we used a magnitude-limited catalogue. This means that the radial distribution of galaxies must be weighted in order to account for the decrease in the number density of galaxies with increasing redshift and produce an isotropic distribution. We used the formula provided by \citet{Efstathiou01} to model the change in the mean number of galaxies as a function of their redshift as:

\begin{equation}\label{eq:sdss_number_density}
N(z) \; dz = A \; z^2  \; \psi(z) dz
\end{equation}

\noindent where A is a normalization factor that depends on the density of galaxies, and $\psi(z)$ is the \textit{selection function}:

\begin{equation}
\psi(z) = e^{-(z/z_r )^{\beta}}
\end{equation}

\noindent where $z_r$ is the characteristic redshift of the distribution and $\beta$ encodes the slope of the curve. Figure \ref{fig:SDSS_selection_function} shows the distribution of galaxies in the SDSS with redshift and the best fit to equation \ref{eq:sdss_number_density}.

\subsubsection{Mean inter-galaxy separation}

Figure \ref{fig:SDSS_mean_separation} shows the mean inter-galaxy separation as a function of redshift. This plot does not account for small-scale clustering and so the values are average over the volume. We can estimate the real inter-galaxy distance by considering that galaxies are contained in approximately $\%10$ of the volume, the line in Fig. \ref{fig:SDSS_mean_separation} must be divided by $\sqrt 10 \sim 3$ giving a mean inter-galaxy separation of 0.8-1.6 Mpc on the redshift range $z = 0-0.1\;$.

%
\begin{figure*}
  \begin{center}
       \includegraphics[width=0.99\textwidth,angle=0.0]{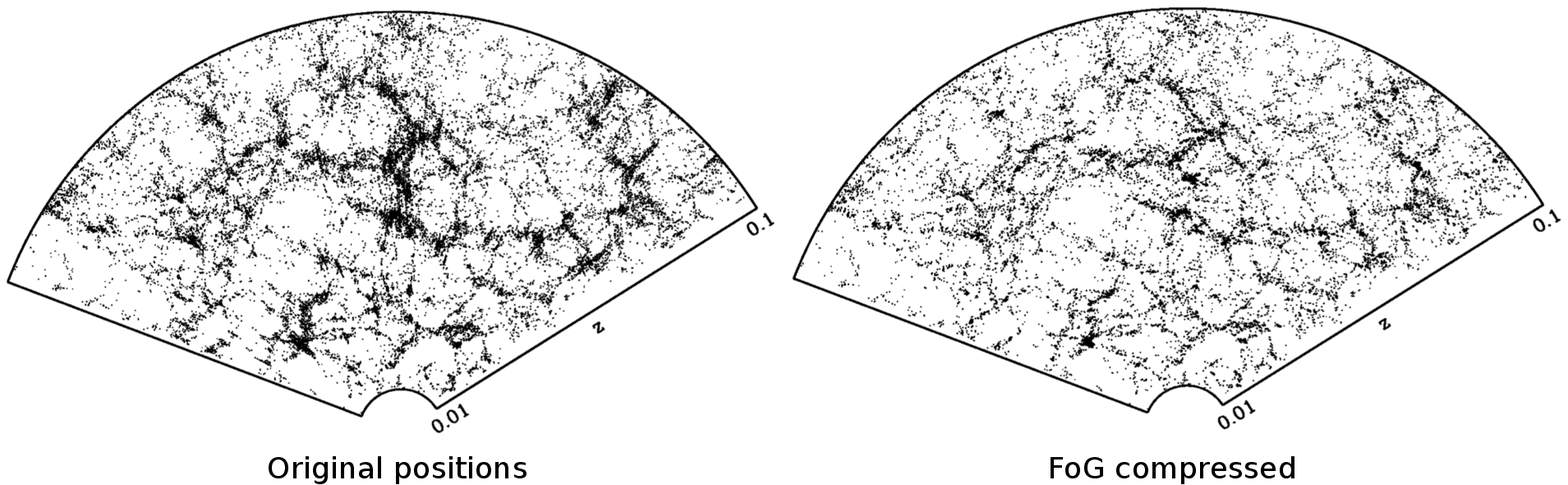}
  \end{center}
  \vspace{-0.5cm}
  \caption{\small {\bf Finger of God compression.} Fingers of God removal in a slice of 6 degrees of thickness in declination. From the original galaxy distribution we extract the Fingers of God (middle left) and compress them on the LoS direction (middle right). The final distribution without FoGs is shown in the lower slice.}
  \label{fig:FoG_compression}
\end{figure*}

\section{Finger of God compression} \label{sec:fog_compression}

The peculiar velocities in dense non-linear regions such as clusters produce the so-called Finger of God (FoG) and are an important source of contamination for both the density estimation and the Cosmic Web identification. We use a FoG compression algorithm following the approach of \citet{Tegmark04} and identify elongated groups using a Friends of Friends group finder where galaxies are recursively linked to other galaxies within a linking volume defined by the linking parameter $b$:

\begin{equation}
b = \left ( \frac{4}{3} \pi  \; \bar{n}_{gal} \; r^3_{link} \right )^{1/3}
\end{equation}

\noindent where $\bar{n}_{gal}$ is the mean number density of galaxies and $r^3_{link}$ is the linking length. In redshift space peculiar velocities in clusters given them an elongated shape along of the LoS. Following \citet{Huchra82} we decompose the distance between galaxies $i$ and $j$ into LoS distance:

\begin{equation}
D_{\parallel ,i,j} = (c/H_0) (z_i+z_j) \sin (\theta_{i,j} / 2)
\end{equation}

\noindent and transversal distance:

\begin{equation}
D_{\perp ,i,j} = (c/H_0) \mid z_i+z_j \mid
\end{equation}

\noindent where $z_i,z_j$ are the redshifts to each galaxy and $\theta_{i,j}$ os their angular separation. Two galaxies are linked if the two following conditions are satisfied:

\begin{equation}
D_{\parallel ,i,j} \le  b_{\parallel} \bar{n}_{gal}(z)^{-1/3} 
\end{equation}

\noindent and 

\begin{equation}
D_{\perp ,i,j} \le  b_{\perp} \bar{n}_{gal}(z)^{-1/3} 
\end{equation}

\noindent where $\bar{n}_{gal}(z)$ is the mean number density of galaxies as function of redshift and $b_{\parallel}, b_{\perp} $ are the projected linking lengths. A typical relation between LoS and transverse linking parameters is $b_{\parallel} = 8\; b_{\perp}$ \citep{Tegmark04, Berlind06}. The elongated groups are then \textit{compressed} such that their dispersion along the LoS is equal to the dispersion in the plane of the sky. Figure \ref{fig:FoG_compression} shows the effect of the FoG compression on the galaxy distribution and on the underlying density field. One drawback of the FoG compression is that systematically compresses filaments oriented along the LoS \citep{Jones10}.

%

\section{Efficient Watershed-Based Distance transform Computation} \label{app:distance_transform}

There are several algorithms for computing distance transforms using the sampling grid itself as a measure of metric. However, for an euclidean metric this operation is $O(N^2)$, which makes computing euclidean distance transforms in large 3D grids computationally expensive and even prohibitive. However, we can take advantage of the space-partitioning property of the watershed transform and compute distance transforms inside each watershed region independently (see also \citet{Steinberg14} for a similar application of the Voronoi tessellation for N-body computations). By doing so we reduce the number of computations from $O(N^2)$ to $O((N/M)^2)$ where M is the number of watershed regions in the volume. For cosmological volumes, such as the SDSS survey we analyze here, the large number of watershed regions (voids) makes the brute-force computation of euclidean distances feasible. Since each watershed region can be treated independently this optimization is trivial to parallelize. 
In addition to the optimization described above, the brute-force distance transform computation is done on the GPU using a CUDA code adapted from the GPU-enabled gravitational code described in \citet{Nguyen07}. If the number of voxels inside a given watershed region exceeds the available memory in the GPU we split the watershed region into smaller blocks and process them independently. Finally the partial computations are combined to find the minimum distances required in equation \ref{eq:distance_transform}.

\end{document}